\documentclass{article} % For LaTeX2e
\usepackage[preprint]{neurips_2024}
\usepackage{times}

\usepackage{amsmath,amsfonts,bm}

\def\eqref#1{equation~\ref{#1}}
\def\1{\bm{1}}

\DeclareMathAlphabet{\mathsfit}{\encodingdefault}{\sfdefault}{m}{sl}
\SetMathAlphabet{\mathsfit}{bold}{\encodingdefault}{\sfdefault}{bx}{n}

\usepackage{hyperref}
\usepackage{url}
\usepackage{xspace}
\usepackage{graphicx}
\usepackage{algorithm}
\usepackage{amssymb}
\usepackage{algpseudocode}

\newcommand*{\res}[1]{\num[round-mode=places,round-precision=2]{#1}}

\usepackage{listings}
\usepackage{hyperref}       % hyperlinks
\usepackage{url}            % simple URL typesetting
\usepackage{booktabs}       % professional-quality tables
\usepackage{amsfonts}       % blackboard math symbols
\usepackage{nicefrac}       % compact symbols for 1/2, etc.
\usepackage{microtype}      % microtypography
\usepackage{xcolor}         % colors
\usepackage{xspace}
\usepackage{hyperref}
\usepackage{url}
\usepackage{xspace}
\usepackage{multirow}
\usepackage{graphicx}
\usepackage{caption}
\usepackage{subcaption}
\usepackage{pifont}
\usepackage{wrapfig}
\usepackage{amsmath, amssymb, amsthm}
\usepackage[capitalise]{cleveref}
\usepackage{etoolbox}
\usepackage{booktabs}
\usepackage{acronym}
\usepackage{siunitx}
\usepackage{bbm}

\newif\ifdraft
\draftfalse

\ifdraft

    \newcommand{\bsc}[1]{\textcolor{blue}{[BS: #1]}}
    \newcommand{\gl}[1]{{\color{red}[G: #1]}}

\else
    \newcommand{\bsc}[1]{}
    \newcommand{\gl}[1]{}

\fi

\title{High Fidelity Text-Guided Music Editing via Single-Stage Flow Matching}

\author{%
  \textbf{Gael Le Lan}% \thanks{Equal contribution. Corresponding authors: \texttt{\{glelan\}@meta.com}} \\
  \quad
  \textbf{Bowen Shi}
  \quad
  \textbf{Zhaoheng Ni}
  \quad
  \textbf{Sidd Srinivasan}\\
  \quad
  \textbf{Anurag Kumar}
  \quad
  \textbf{Brian Ellis}
  \quad
  \textbf{David Kant}
  \quad
  \textbf{Varun Nagaraja}
  \quad
  \textbf{Ernie Chang}\\
  \quad
  \textbf{Wei-Ning Hsu}
  \quad
  \textbf{Yangyang Shi}
  \quad
  \textbf{Vikas Chandra}\\
  AI at Meta\\
  \texttt{\{glelan,bshi\}@meta.com}
}

\newcommand{\proposed}{\textsc{MelodyFlow}\xspace}

\newcommand{\musicgen}{\textsc{MusicGen}\xspace}
\newcommand{\audioldm}{\textsc{AudioLDM}\xspace}
\newcommand{\stableaudio}{\textsc{Stable-Audio}\xspace}
\newcommand{\magnet}{\textsc{MAGNeT}\xspace}
\newcommand{\renoise}{\textsc{ReNoise}\xspace}

\begin{document}

\maketitle

\begin{abstract}
We introduce \textsc{MelodyFlow}, an efficient text-controllable high-fidelity music generation and editing model.
It operates on continuous latent representations from a low frame rate 48 kHz stereo variational auto encoder codec.
% that eliminates the information loss drawback of discrete representations.
Based on a diffusion transformer architecture trained on a flow-matching objective the model can edit diverse high quality stereo samples of variable duration, with simple text descriptions.
We adapt the \textsc{ReNoise} latent inversion method to flow matching and compare it with the original implementation and naive denoising diffusion implicit model (DDIM) inversion on a variety of music editing prompts.
Our results indicate that our latent inversion outperforms both \textsc{ReNoise} and DDIM for zero-shot test-time text-guided editing on several objective metrics.
Subjective evaluations exhibit a substantial improvement over previous state of the art for music editing. Code and model weights will be publicly made available. Samples are available at \url{https://melodyflow.github.io}.

% We also explore a new regularized latent inversion method for zero-shot test-time text-guided editing and demonstrate its superior performance over naive denoising diffusion implicit model (DDIM) inversion for variety of music editing prompts.
% Evaluations are conducted on both objective and subjective metrics and demonstrate that the proposed model is not only competitive to the evaluated baselines on a standard text-to-music benchmark - quality and efficiency-wise - but also outperforms previous state of the art for music editing when combined with our proposed latent inversion. Samples are available at \url{https://melodyflow.github.io}.
\end{abstract}

% We introduce a simple and fast text-controllable high-fidelity music generation and editing model.
% It operates on sequences of continuous latent representations from a quantizer-free Variational AutoEncoder (VAE) 48 kHz stereo codec that eliminates the information loss drawback of discrete representations.
% Based on a diffusion transformer architecture trained on a flow-matching objective, the model can generate diverse high quality stereo samples, with simple text descriptions.
% We also explore text-guided style transfer and editing and demonstrate the impressive performance of the model for a variety of music editing tasks.
% Evaluations are conducted on both objective and subjective metrics and demonstrate that the proposed model is not only superior to the evaluated baseline on a standard text-to-music benchmark but also benefits from impressive style-transfer capabilities that are inherent to the model design.
% Model implementation and weights will be publicly made available to foster further research on music editing.

\section{Introduction}

Text-conditioned music generation has made tremendous progress in the past two years \citep{schneider2023mo,huang2023noise2music,agostinelli2023musiclm,copet2024simple,ziv2023masked,liu2023audioldm2,li2023jen,prajwalmusicflow}.
% The dominant approach involves representing audio as a sequence of compressed discrete or continuous representations and training a generative model on top of it.
The prevailing method for audio representation involves compressing the waveform into a series of discrete or continuous tokens, and then training a generative model on top of those.
Two dominant generative model architectures have emerged, one based on autoregressive Language Models (LMs) \citep{agostinelli2023musiclm,copet2024simple}, the other on diffusion \citep{schneider2023mo,huang2023noise2music,liu2023audioldm2,li2023jen,prajwalmusicflow}.
A third method sometimes referred to as discrete diffusion relies on non-autoregressive masked token prediction \citep{ziv2023masked,garcia2023vampnet}.
The target level of audio fidelity depends on the models and some have already successfully generated 44.1 kHz or high stereo signals \citep{schneider2023mo,li2023jen,evans2024fast}.

% However no such work has really discussed the implications of aiming for higher fidelity or stereo and what performance compromise it implies from the latent representation and generation model perspective.

The increasing popularity of diffusion models in computer vision has led to the emergence of a new area of research focused on text-controlled audio editing \citep{wang2023audit,lin2024arrange,garcia2023vampnet,wu2023music,novack2024ditto,zhang2024musicmagus,manor2024zero}.
The sound design process often involves multiple iterations, and using efficient editing methods is a key approach to achieving this effectively.
Music editing encompasses a wide range of tasks, including but not limited to: inpainting/outpainting, looping, instrument or genre swapping, vocals removal, lyrics editing, tempo control, and recording conditions modification (e.g. from studio quality to a concert setting).
Recent works have addressed some of these tasks using specialized models \citep{wang2023audit,garcia2023vampnet,lin2024arrange,wu2023music,copet2024simple} or zero-shot editing methods from the computer vision domain, which are exclusive to diffusion models \citep{novack2024ditto,zhang2024musicmagus,manor2024zero}.
Despite recent efforts, no approach has yet shown the ability to perform high-fidelity generic style transfer across various music editing tasks. This limitation can be attributed to several factors, including insufficient high-quality data, inadequate foundational music generation models, and design choices that fail to generalize effectively to diverse editing tasks.
% However, none of these have demonstrated high fidelity generic style transfer capabilities, due to either lack of high quality data or foundational music generation model, or design choices that do not generalise to any kind of editing task.
% The question of inference speed is key for creatives and the music domain is particularly challenging due to the high fidelity (48 kHz stereo) requirement in the sound design process.
Inference speed is crucial for creatives, and the music domain presents a unique challenge due to the high-fidelity (48 kHz stereo) requirement in the sound design process.
Recently \citet{lipman2022flow} proposed the Flow Matching (FM) generative modeling formulation, which involves constructing optimal transport paths between data and noise samples.
Flow Matching (FM) offers a more robust and stable approach to training diffusion models, with the added benefit of faster inference.
This method has been successfully applied to train foundational speech \citep{le2024voicebox} and audio \citep{vyas2023audiobox} generative models.
For the music domain \citet{prajwalmusicflow} utilized a two-stage FM model for text-guided music generation, where the first stage generates semantic features and the second stage generates acoustic features.
% cascading semantic and acoustic features generation.
% It is a more robust and stable alternative for training diffusion models with notably faster inference and was successfully applied to train foundational speech \citep{le2024voicebox} and audio \citep{vyas2023audiobox} generative models.

In this work we present \proposed, a single-stage text-conditioned FM model designed for instrumental music generation and editing.
% It is able to generate and edit high fidelity stereo samples using text descriptions.
The model operates on continuous representations of a low frame rate Variational Audio Encoder (VAE) codec.  % , which is crucial for handling high-fidelity stereo generation efficiently compared to models operating on discrete neural codecs.
% few key design choices our model is also notably efficient.
% Indeed despite its smaller size the generated samples quality match previously 5 times bigger LM-based models.
% Our evaluations show that the proposed model is competitive with evaluated baselines on text-to-music generation while being remarkably efficient, with a latency of less than 10 seconds for a 30 seconds sample generation (64 inference steps).
Additionally, thanks to the versatility of FM, \proposed is compatible with any zero-shot test-time editing method such as DDIM inversion \citep{song2020denoising} or ReNoise \citep{garibi2024renoise}.
We enhance the editability of the FM inversion by adapting the latent inversion of \citet{garibi2024renoise} to the FM formulation.
% process with a regularized latent inversion method inspired by that of \citep{garibi2024renoise} for diffusion-based image editing.
Both our objective and subjective evaluations on music editing indicate that \proposed can support a diversity of editing tasks on real songs without any finetuning, achieving fast music editing with remarkable consistency, text-adherence and minimal quality loss compared with original samples.
%strongly outperforming previous state of the art on the task.
In addition we conduct an ablation study on the importance of the key design choices on the overall model quality/efficiency trade off.

\textbf{Our contributions:} (i) We introduce the first of its kind single-stage text-to-music FM model to generate and edit 48 kHz stereo samples of up to 30 seconds, with enhancements in both the audio latent representation and generative model, striking a better balance between quality and efficiency. (ii) We explore a novel regularized FM inversion method capable of performing faithful zero-shot test-time text-guided editing on various axes while maintaining coherence with the original sample. (iii) We publicly release the code and model weights to foster research on music editing.\bsc{One contribution is to improved design of FM for audio generation, including logit-normal sampling, VAE-codec, stereo audio}

% (iii) We conduct comprehensive objective and human evaluations on the key design choices underlying our approach.

% \textbf{Our contribution:} (i) We introduce an efficient single-stage flow matching model to generate high quality stereo music at 48 kHz thanks to a continuous VAE codec that operates at much lower frame rate than previous discrete representations. (ii) We demonstrate that the model is able to perform faithful text-guided editing on a wide range of axes while keeping the generated audio coherent with the original sample. (iii) We provide extensive objective and human evaluations on the key design choices behind our method.

\section{Method}

\begin{figure}[t!]
     \centering
     \includegraphics[width=\textwidth]{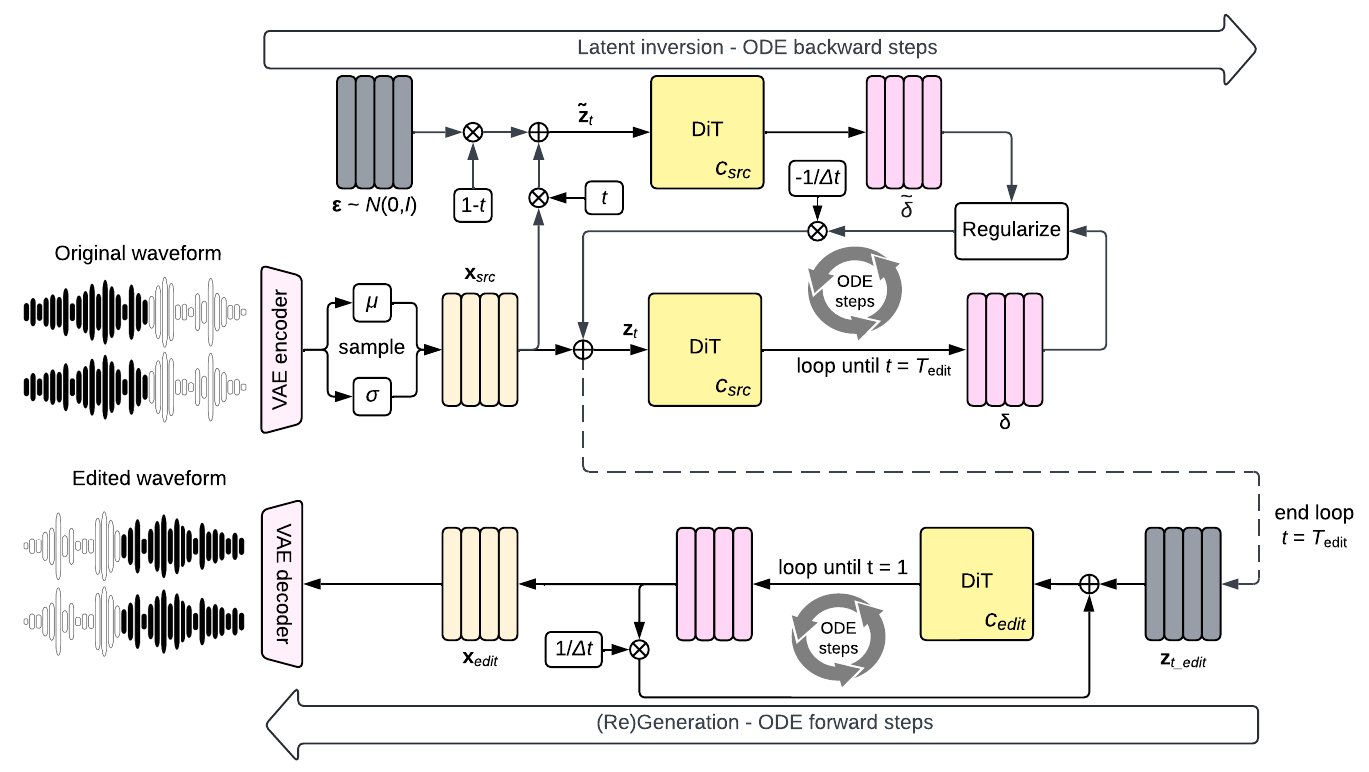}
     % \vspace{-0.2cm}
     \caption{Overview of the \proposed editing process.
     A waveform is encoded into $\mathbf{x}_{src}$ before being fed to the ODE solver. 
     Step-by-step, the DiT predicts the velocity $\delta$ from data to noise, while being regularized against the prediction of an artificially constructed $\mathbf{\tilde{z}}_t$ so as to enhance editability.
     Once the target inversion flow step $T_{edit}$ has been reached, the model is used in the classic generation setting (bottom of the Figure, from right to left), except that the starting latent $\mathbf{z}_{t_{edit}}$ has been estimated so as to achieve better editability and consistency with the source waveform.}
     % , while offering great editability.}
     % is the target inversion timestep. Given $S$ ODE solver steps, the inversion step size is given by $\Delta{t} = (1 - T_{edit})/{S}$. The editing prompt $c_{edit}$ is not used during inversion.}
     \label{fig:general}
     % \vspace{-0.2cm}
\end{figure}

\proposed combines a continuous audio codec, a text-conditioned Diffusion Transformer (DiT) FM model and a regularized latent inversion method.
The model can perform text-guided editing of real or generated audio samples.
The overall editing process is depicted in the Figure \ref{fig:general}.
% A neural quantizer-free audio codec with a VAE bottleneck extracts a sequence of latent audio representations that are fed to the DiT that predict the latent inversion (or generation) trajectory.
% operates on sequences of latent audio representations from a neural quantizer-free audio codec with a VAE bottleneck.
% Combined with our regularized latent inversion, 
% The appendix \ref{apx:details} shares additional details regarding the model key components (codec, generative model and latent inversion implementation).
% can be found in the appendix \ref{apx:details}.
% The quality of music editing depends on the codec, FM model and the latent inversion method, while we know speed is a crucial factor in the sound design creative process.

% keeping in mind that speed is a crucial factor in the sound design creative process
% The quality of music editing depends on the codec, FM model and the latent inversion method, while we know speed is a crucial factor in the sound design creative process.

\subsection{Latent audio representation}

Our codec derives from EnCodec \citep{defossez2022high} with additional features from the Descript Audio Codec (DAC) \citep{kumar2024high} (snake activations, band-wise STFT discriminators) and \citet{evans2024fast} (KL-regularized bottleneck, perceptual weighting).
A convolutional auto-encoder encodes the waveform into a sequence of latent bottleneck representations, its frame rate function of the convolution strides.
% Given a fixed target latent dimension, the lower the sequence frame rate, the higher the compression ratio.
Audio fidelity is enforced by multi-scale STFT reconstruction losses complemented by the sum and difference STFT loss for stereo support \citep{steinmetz2020automatic}.

\subsection{Conditional flow matching model}
Given an audio sample $\mathbf{a}\in\mathbb{R}^{D\times f_s}$, a sequence $\mathbf{x}\in\mathbb{R}^{L\times d}$ of latent representations is extracted by the neural codec.
FM models the optimal transport paths that map a sequence $\mathbf{\epsilon}\in\mathbb{R}^{L\times d}\sim\mathcal{N}(0,I)$ to $\mathbf{x}$ trough a linear transformation - function of the flow step $t$ - following equation \ref{eq:fm_mixture}.
\[\mathbf{z}_t=t\mathbf{x} + (1-t)\mathbf{\epsilon}, t \in [0,1]\label{eq:fm_mixture}\]
During training, $t$ is randomly sampled and the DiT $\Theta$ is trained to estimate $d\mathbf{z}_t/dt$ conditioned on $t$ and a text description $c$.
\[d\mathbf{z}_t/dt=v_{\Theta}(\mathbf{z}_t,t,c)=\mathbf{x}-\mathbf{\epsilon}\]
By design, after training, the model can be used with any ODE solver to estimate $\mathbf{x}=\mathbf{z_1}$ given $\mathbf{\epsilon}=\mathbf{z_0}$ (and vice versa), and a text description.
The text-to-music inference happens as such: starting from a random noise vector $\mathbf{\epsilon}\in\mathbb{R}^{L\times d}\sim\mathcal{N}(0,I)$ and a text description $c$ of the expected audio the ODE solver is run from $t=0$ to $t=1$ to estimate the most likely sequence of latents $\mathbf{x}_{generated}$.
\[\mathbf{x}_{generated}=\textbf{ODE}_{0\xrightarrow{}1}(\epsilon,c)\]
After the latents have been estimated they are fed to the codec decoder to materialize the waveform.
% Flow matching models optimal transport paths that map a sequence $\mathbf{\epsilon}\in\mathbb{R}^{L\times d}\sim\mathcal{N}(0,I)$ to $\mathbf{x}$ trough a continuous transformation of probability densities.
% The flow is a time-dependent mapping $\phi_t:[0,1]\times\mathbb{R}^d\longrightarrow\mathbb{R}^d$ defined by the following ordinary differential equation (ODE).
% \[d\mathbf{z}_t=v_{\Theta}(\mathbf{z}_t,t)dt\]
% \gael{reformulate the math + minibatch coupling}
%$\Theta$ is a neural network that parameterizes the velocity $v$.
%The flow matching objective is to estimate the derivative $v$ from the input mixture of signal and noise defined hereafter.
%\[\mathbf{z}_t=t\mathbf{x} + (1-t)\mathbf{\epsilon}\]
%Training the model consists in minimizing the following loss over $(t,\mathbf{x},\mathbf{\epsilon})$.
%\[
%\mathcal{L}_{CFM} = \mathbb{E}_{t,p(\mathbf{x}|\epsilon),q(\epsilon)}||v_t(\mathbf{x},\Theta) - u_t(x|\epsilon)||^2
%\]
\citet{kingma2024understanding} show that the flow step sampling density during training plays an important role in model performance.
In our implementation $t$ is sampled from a logit-normal distribution \citep{karras2022elucidating,esser2024scaling}.
% Logit-normal sampling was indeed originally proposed in \citep{karras2022elucidating} for v-prediction and also showed promising results in \citep{esser2024scaling}, both for image generation.

% Indeed considering the flow matching problem formulation, the signal-to-noise ratio in $\mathbf{z_t}$ tends to $\pm\infty$ when $t$ tends to either 0 or 1.
% Thus it does not make sense for the model to spend too much capacity where the derivative is almost impossible to estimate.
% By design, after training, the model is directly compatible with any ODE solver.
% The text-to-music inference happens as such: starting from a random noise vector $\mathbf{\epsilon}\in\mathbb{R}^{L\times d}\sim\mathcal{N}(0,I)$ and a text description of the expected output $c$ (materializing the model input at $t=0$) the ODE solver is run from $t=0$ to $t=1$ to estimate the most likely sequence of latents $\mathbf{x}_{generated}$.
% \[\mathbf{x}_{\text{generated}}=\textbf{ODE}_{0\xleftarrow{}1}(\epsilon,c)\]
% Classifier-free guidance is employed.
%After the latents have been estimated they are fed to the codec decoder to materialize the waveform.

\subsection{Text-guided editing through latent inversion}
Due to the bijective nature of the FM formulation (where given a text condition, each latent sequence is mapped to a single noise vector), the model is compatible with existing latent inversion methods such as DDIM inversion \citep{song2020denoising}.
Given the latent representation $\mathbf{x}_{src}$ of an existing audio with an optional accompanying caption $c\in\{\varnothing,c_{src}\}$, the model can estimate its corresponding noise (or intermediate) representation $\mathbf{z}_{t_{edit}}=\textbf{ODE}_{t_{edit}\xleftarrow{}1}(\mathbf{x}_{src},c)$ by running the ODE solver in the backward direction until an intermediary time step $t_{edit}$ (top of the Figure \ref{fig:general}).
% It is important to note that neither $\mathbf{x}_{src}$ nor $c_{src}$ have to be human-generated.
% $\mathbf{x}_{src}$ could be the output of a preceding text-to-music or music editing process and $c_{src}$ could result from a music captioning model.
Given the intermediary representation $\mathbf{z}_{t_{edit}}$, the ODE forward process can be conditioned on a new text description $c_{edit}$ that materialises the editing prompt: $\mathbf{x}_{edit}=\textbf{ODE}_{t_{edit}\xrightarrow{}1}(\mathbf{z}_{t_{edit}},c_{edit})$.
A good inversion process should accurately reconstruct the input when $c_{edit}=c_{src}$, as shown in equation \ref{eq:ode_inv}.
\[
\mathbf{x}_{edit}=\textbf{ODE}_{t_{edit}\xrightarrow{}1}(\textbf{ODE}_{t_{edit}\xleftarrow{}1}(\mathbf{x}_{src},c\in\{\varnothing,c_{src}\}),c_{src}) \approx \mathbf{x}_{src}
\label{eq:ode_inv}
\]
In such case when swapping $c_{src}$ for $c_{edit}$ in the $t_{edit}\xrightarrow{}1$ forward direction, the expectation is for the generated audio to preserve some consistency with the source while being faithful to the prompt.
However in practice it was observed by \citet{mokady2023null} that DDIM inversion suffers from poor editability due to the classifier free guidance.

% , while we know speed is a crucial factor in the sound design creative process. Some of our design choices are critical in striking a good balance between quality and efficiency.

\bsc{The following two sections aren't related to editing and better moved to a separate section on improving FM for TTM}\gl{done}

\subsection{Regularized latent inversion}%\gael{define $L_{edit}$}
% \citep{huberman2023edit} recently observed that DDPM model predictions in subsequent solver steps are usually correlated, which does not correspond to the theoretical assumption of independence of the predicted noise maps.
% They manually design a latent inversion method that better mimics uncorrelated noise prediction assumptions.
% This leads to better editability and perfect reconstruction accuracy.
% \citep{garibi2024renoise} recently observed a similar intuition for DDIM inversion.
% They observe that the denoising trajectory is not easily reversible.

%, where the model prediction is not noise but velocity.
% This mechanism refines the approximation of a predicted point along the forward diffusion trajectory, by iteratively applying the pretrained diffusion model and averaging its predictions.
% In practice DDIM inversion suffers from poor editability and reversibility.
Even though FM consists in estimating straight trajectories, in practice those are never completely straight and the edited samples do not preserve enough consistency with the source.

% DDIM inversion suffers from poor editability and reversibility.

\begin{enumerate}
\item The distribution of predicted velocities tends to shift away from that of training due to the classifier free guidance \citep{mokady2023null}, which can lead to divergence of the inversion trajectory. This was observed by \citet{parmar2023zero} with $\epsilon$-prediction, which they address by adding an autocorrelation regularization during inversion to preserve the statistical properties of the predictions.
% One way to overcome this is to regularize them via gradient descent.
\item Any pair of successive $(\mathbf{z}_{t},\mathbf{z}_{t - \Delta t})$ along the inversion path usually has estimated velocities $v_{\Theta}(\mathbf{z}_{t}, t, c) \neq v_{\Theta}(\mathbf{z}_{t - \Delta t}, t - \Delta t, c)$, which affects reversibility (hence the consistency with the source sample).
Building a fully reversible inversion path requires estimating $\mathbf{z'}_{t - \Delta t}$ such that $v_{\Theta}(\mathbf{z}_{t}, t, c) \approx v_{\Theta}(\mathbf{z'}_{t - \Delta t}, t - \Delta t, c)$, for example following \citet{garibi2024renoise}.

% Given a manually constructed $\mathbf{\tilde{z}}_{t}=\mathbf{x}t + \epsilon(1-t)$ and a real $\mathbf{z}_{t}$ along the inversion trajectory, and corresponding predictions $v_{\Theta}(\mathbf{z}_{t}, t, c)$ and $\tilde{v}_{\Theta}(\mathbf{\tilde{z}}_{t}, t, c)$. We arrange model predictions in 4x4 patches and compute the average Kullback-Leibler (KL) divergence $\mathcal{L}_{patchKL}$ between corresponding patches.
% we can compute the Kullback-Leibler (KL) divergence between predictions $v_{\Theta}(\mathbf{z}_{t}, t, c)$ and $\tilde{v}_{\Theta}(\mathbf{\tilde{z}}_{t}, t, c)$ and regularize $v_{\Theta}$ via gradient descent. We employ the same patch-wise KL divergence computation as ReNoise.
\end{enumerate}
% The main difference between our method and ReNoise is that given the FM formulation we only enforce the model prediction regularization, but neither de-correlation nor noise correction steps.
% In appendix \ref{apx:add_exp} we compare our approach with ReNoise by reformulating the FM prediction into a noise prediction task. 

% One way to construct a fully reversible inversion path is by enforcing reversibility for any pair of successive $(\mathbf{z}_{t_1},\mathbf{z}_{t_2})$ along the inversion path.
% \[
% \mathbf{z}_{t_2} = \mathbf{z}_{t_1} + (t_1-t_2) v_{\Theta}(\mathbf{z}_{t_1}, t_1, c)
% \label{eq:renoise_inversion_1}
% \]
% \[
% \mathbf{z}_{t_1} = \mathbf{z}_{t_2} - (t_1-t_2) v_{\Theta}(\mathbf{z}_{t_2}, t_2, c)
% \label{eq:renoise_inversion_2}
% \]
% \[
% v_{\Theta}(\mathbf{z}_{t_1}, t_1, c) = v_{\Theta}(\mathbf{z}_{t_2}, t_2, c)
% \label{eq:renoise_inversion_3}
% \]

% proposed a latent inversion method that rely on regularizing the DDPM inversion path for better reconstruction accuracy and editability. Their observation is that the model predictions do not follow the expected noise distribution in diffusion models.
% \citep{garibi2024renoise} follows a similar intuition for DDIM inversion.

\renoise \citep{garibi2024renoise} addresses those two problems by combining both $\epsilon$-prediction regularization and reversible inversion trajectory estimation.
% However it was defined in the context of $\epsilon$-prediction (in such case applying an autocorrelation regularization is natural), and the FM formulation consists in $v$-prediction (where such regularization would be detrimental considering the presence of signal in the prediction).
Applying \renoise to FM requires either (1) reformulating FM as $\epsilon$-prediction or (2) adapting the regularization mechanism.
Indeed since our FM model predicts the velocity $v_{\Theta}(\mathbf{z}_t,t,c)=\mathbf{x}-\mathbf{\epsilon}$ and \renoise operates on noise predictions, applying \renoise to FM (1) requires subtracting the source latent $\mathbf{x}_{src}$ from $v_{\Theta}$ to try and isolate and regularize $\mathbf{\epsilon}$ directly.
However in such setting the inversion diverges when conditioning on text ($c_{src}$) and using CFG (appendix \ref{apx:cfg}), likely due to $v_{\Theta}(\mathbf{z}_t,t,c)-\mathbf{x}_{src}$ not properly removing the signal component of $\mathbf{x} - \epsilon$ at lower flow steps.
To prevent this behavior we propose to (2) directly regularize the FM prediction using only the KL regularization from \citet{garibi2024renoise}.
An thorough comparison between the considered approaches can be found in the sections \ref{exp:inv_methods}, \ref{exp:inv_target} and \ref{exp:inversion}.
% More details about our choice can be found in appendix \ref{apx:add_exp}, along with an experimental comparison between the considered approaches.
% We also provide a comparative analysis with DDIM inversion in sections \ref{exp:inversion} and \ref{exp:tedit}.

The Algorithm \ref{alg:cap} details our proposed inversion. Each iteration consists in estimating a reversible inversion point $\mathbf{z}_{t-\Delta t}$ from a source point $\mathbf{z}_{t}$ such that $v_{\Theta}(\mathbf{z}_{t-\Delta t}, t-\Delta t, c) \approx v_{\Theta}(\mathbf{z}_{t}, t, c)$. In such case the jump from $\mathbf{z}_{t}$ to $\mathbf{z}_{t-\Delta t}$ is considered reversible. This is done iteratively in $K$ steps following the convergence property of \citet{garibi2024renoise}.
During each of those steps, the model prediction is regularized against the prediction of an artifically constructed $\mathbf{\tilde{z}}_{t-\Delta{t}}$ (also shown in the Figure \ref{fig:general}).

% Each step consists in predicting the next jump $\delta$

% achieve by averaging the regularized predictions from $K$ subsequent steps.
% requires $K$ estimations to obtain a reversible point (which is average from the .
\bsc{Better to elaborate more on this part, since this is the key contribution of this paper. }

\begin{algorithm}
\caption{Proposed regularized FM inversion}\label{alg:cap}
\begin{algorithmic}
\Require Sequence of audio latents $\mathbf{x}$. Number of ODE backward steps $S$. Source text description $c\in\{\varnothing,c_{src}\}$. K regularization steps with weights $\{w_k\}_{k=1}^{K}$, KL regularization weight $\lambda_{KL}$. 
\Ensure A noisy latent $\mathbf{z}_{T_{edit}}$ such that $\textbf{ODE}_{T_{edit}\xrightarrow{}1}(\mathbf{z}_{T_{edit}},c_{src})\approx\mathbf{x}$.

\State $\Delta{t} \gets (1 - T_{edit})/{S}$
% $\mathbf{z}_1=\mathbf{x}$

\For{$t=1,1-\Delta{t},\dots,T_{edit}+\Delta{t}$}
\State $\mathbf{z}^{(0)}_{t-\Delta{t}} \gets \mathbf{z}_{t}$
%\State $\mathbf{z}^{(avg)}_{t-\Delta{t}} \gets 0$
\For{$k=1,\dots,K$}
\State $\delta \gets v_{\Theta}(\mathbf{z}^{(k-1)}_{t-\Delta{t}}, t-\Delta{t}, c)$
\If{$w_k > 0$}
\State sample $\epsilon \sim \mathcal{N}(0, I)$
\State $\mathbf{\tilde{z}}^{(k-1)}_{t-\Delta{t}} \gets \mathbf{x}(t-\Delta{t}) + \epsilon(1-(t-\Delta{t}))$
\State $\tilde{\delta} \gets v_{\Theta}(\mathbf{\tilde{z}}^{(k-1)}_{t-\Delta{t}}, t-\Delta{t}, c)$
\State $\delta \gets \delta - \lambda_{KL}\nabla_\delta\mathcal{L}_{patchKL}(\delta, \tilde{\delta})$
\EndIf
\State $\mathbf{z}^{(k)}_{t-\Delta{t}} \gets \mathbf{z}_{t} - \delta\Delta{t}$
\EndFor
\State $\mathbf{z}_{t-\Delta{t}} \gets \frac{\sum_{k=1}^{K} w_k \mathbf{z}^{(k)}_{t-\Delta{t}}}{\sum_{k=1}^{K} w_k}$
%\State $\mathbf{z'}_{t-\Delta{t}} \gets \mathbf{z}^{(avg)}_{t-\Delta{t}}$
% $\mathbf{z}_{t+\Delta{t}}=\textbf{ODE}_{t+\Delta{t}\xleftarrow{}t}(\mathbf{z}_{t},c)$
\EndFor

\State \Return $\mathbf{z}_{T_{edit}}$

\end{algorithmic}
\end{algorithm}

\subsection{Improving Flow Matching for text-to-music generation}

% The quality of music editing depends on the three components: codec, FM model and latent inversion method.
% Improvements on the codec and FM model not only benefit music editing but also text-to-music generation.
% Given that speed is a crucial factor in the sound design creative process, the following design choices are critical in striking a good balance between quality and efficiency.
% , which we highlight hereafter.

\subsubsection{Codec bottleneck}

Recently \citet{prajwalmusicflow} trained a two-stage music FM model on continuous latent representations, but both the semantic and acoustic latent representations where trained with a discretization objective (HuBERT semantic features and RVQ-regularized codec).
The concurrent work of \citet{evans2024longform} demonstrated long form music generation capabilities by using a KL-regularized bottleneck in their codec with a temporal downsampling as low as 21.5 Hz.
However none of these works have carefully investigated the influence of the bottleneck regulariser on both music reconstruction and generation performance, all other things being equal.  
% , especially for a similar codec framerate.
% extracted from the encoder of a discrete audio codec (before the RVQ layer), here we completely remove the RVQ layer to optimize for continuous modeling.
% Compared with \citet{vyas2023audiobox} we swap the Residual Vector Quantization (RVQ) layer of the codec with a VAE bottleneck.
% This unlocks higher quality generation at a lower frame rate (e.g. faster inference).
% No prior work has investigated the influence of the VAE bottleneck regulariser on music reconstruction and generation performance.
% Although FM with VAE has already been explored for image generation \citep{esser2024scaling}, our work is the first to investigate the influence of the VAE regulariser on music reconstruction and generation performance.
Indeed \citet{rombach2022high} - a seminal work on VAE for image generation - note that \textit{LDMs trained in VQ-regularized latent spaces achieve better sample quality} than KL-regularized ones. 
Our ablation in section \ref{exp:bottleneck} leads to a different conclusion.
% demonstrates that this conclusion does not apply for audio.
% We show in appendix \ref{exp:codec} that swapping the RVQ by a VAE bottleneck is key to increase the generation model efficiency and the quality of its outputs.
Using a KL-regularizer achieves indeed better music reconstruction and generation performance for a much lower frame rate, which is key for faster inference.
% and scaling to high-fidelity stereo audio.
% (shown in appendix \ref{exp:stereo}).
% This was also observed by the recent work of \citep{evans2024longform}.

\subsubsection{Minibatch coupling}%\gael{reformulate and expand}
\citet{tong2023improving} and \citet{pooladian2023multisample} expanded over prior work on FM modeling by sampling pairs $(\mathbf{x},\mathbf{\epsilon})$ from the joint distribution given the by the optimal transport plan between the data $\mathbf{X} = \{{\mathbf{x}^{(i)}}\}_{i=1}^B$ and noise $\mathbf{E} = \{{\mathbf{\epsilon}^{(i)}}\}_{i=1}^B$ samples within a batch of size $B$. Essentially this translates into running the Hungarian algorithm so as to find the permutation matrix $\mathbf{P}$ that minimizes $||\mathbf{X}-\mathbf{P}\mathbf{E}||_2^2$.
% proposing to align $\mathbf{X} = \{{\mathbf{x}^{(i)}}\}_{i=1}^B$ and $\mathbf{E} = \{{\mathbf{\epsilon}^{(i)}}\}_{i=1}^B$ via optimal transport solving. Essentially this translates into running the Hungarian algorithm so as to a the permutation matrix $\mathbf{P}$ that minimizes $||\mathbf{X}-\mathbf{P}\mathbf{E}||_2^2$.
They demonstrate it results in straighter optimal transport paths during inference (that are closer to the theoretical linear mapping assumption between noise and data samples) and consequently offers better quality-efficiency trade offs. We shed light on the importance of mini-batch coupling in sections \ref{exp:design} and \ref{exp:inversion} where we underline the overall benefit of our FM model design choices on both music generation and editing.
% on its performance.
% in appendix \ref{exp:efficiency}.
% in better image generation quality for a given 

\section{Experimental setup}
\subsection{Model}

\proposed uses a DiT of sizes 400M (small) and 1B (medium) parameters with U-shaped skip connections \cite{Bao_2023_CVPR}.
The model is conditioned via cross attention on a T5 representation \citep{JMLR:v21:20-074} computed from the text description of the music.
The model integrates a specific L-shaped self-attention mask meant to better generalize to different segment lengths during inference (appendix \ref{apx:length}).
The flow step is injected following \citet{hatamizadeh2023diffit}. Minibatch coupling is computed with \verb+torch-linear-assignement+\footnote{\url{https://github.com/ivan-chai/torch-linear-assignment}}.
\proposed-small (resp. \proposed-medium) is trained on latent representation sequences of 32 kHz mono (resp. 48 kHz stereo) segments of 10 (resp. 30) seconds, encoded at 20 Hz frame rate (resp. 25 Hz).
From the codec perspective the only difference between encoding mono or stereo waveform is the number of input (resp. output) channels for the first (resp. last) convolution of the encoder (resp. decoder): 1 for mono and 2 for stereo.
The appendix \ref{apx:stereo} specifically investigates the impact of encoding stereo instead of mono signals on both reconstruction and generation performance.
More details regarding audio representation and FM model implementation and training are provided in the appendix \ref{apx:exp_setup}.

\subsection{Generation and editing}
For text-to-music generation we use the \verb+midpoint+ ODE solver from \verb+torchdiffeq+ with a step size of $0.03125$.
A classifier free guidance (CFG) of $4.0$ is chosen after grid search (appendix \ref{apx:cfg}).
For music editing we use the same configuration for DDIM inversion. For \renoise and \proposed we use a longer step size of 0.04 to account for the additional forward passes induced by the reversible trajectory estimation.
For DDIM inversion this gives a total of 64 inversion and 64 generation steps (e.g. forward passes through the DiT).
For \renoise and \proposed the inversion takes 25 steps (each of them requires 4 iterations for the reversibility estimation) and 25 forward steps, for a total of 125.
% run our regularized inversion method with the \verb+euler+ ODE solver until $T_{edit}=0.04$ with the same classifier free guidance of $5.0$ applied in both ODE directions.
In summary \proposed's inversion is run with $S=25,K=4,w_0=w_1=0,w_2=2,w_3=3$ and $\lambda_{KL}=0.2$.
% This results in a total of $25\times4=100$ model inference steps for trajectory inversion and 25 additional steps for the ODE forward process.
% Those parameters offer a good trade off between speed and quality and were chosen empirically after testing the method interactively on a few samples.
% For regularization we use for regularization  $\lambda_{KL}=0.08$.
% Then the solver is run forwards conditioned on the editing text prompt, using the same classifier free guidance of $5.0$. For music editing ablations we compare DDIM inversion and ReNoise, and use DDPM inversion for diffusion baselines.

\subsection{Datasets}
\paragraph{Training}

Our training dataset is made of 10K high-quality internal music tracks and the ShutterStock and Pond5 music collections with respectively 25K and 365K instrument-only music tracks, totalling into 20k hours.
All datasets consist of full-length music sampled at 48 kHz stereo with metadata composed of a textual description sometimes containing the genre, BPM and key.
Descriptions are curated by removing frequent patterns that are unrelated to the music (such as URLs). 
For 32 kHz mono models the waveform is downsampled and the stereo channels are averaged.

\paragraph{Evaluation}

For the main text-to-music generation results we evaluate \proposed and prior work on the MusicCaps dataset \citep{agostinelli2023musiclm}.
We compute objective metrics for \proposed and report those from previous literature.
Subjective evaluations are conducted on a subset of 198 examples from the genre-balanced set.
For ablations we rely on an in-domain held out evaluation set different from that of \citet{copet2024simple}, made of 8377 tracks.
The same in-domain tracks are used for objective editing evaluations.
Subjective evaluations of edits are run on a subset of 181 higher fidelity samples from our in-domain test set with LLM-assisted designed prompts (more details in appendix \ref{ref:llm_assist}).

\subsection{Metrics}

We evaluate \proposed using both objective and subjective metrics following the evaluation protocol of \citet{kreuk2022audiogen} and \citet{copet2024simple} for generation.
Reported objective metrics are the Fréchet Audio Distance (FAD) \citep{48813} with VGGish embeddings \citep{hershey2017cnn}, the Kullback–Leibler divergence (KLD) with PASST audio encoder \citep{koutini2021efficient} and CLAP\footnote{\url{https://github.com/LAION-AI/CLAP}} cosine similarity \citep{elizalde2023clap}.
For music editing evaluations we compute the average L2 distance between the original and edited latent sequences (LPAPS \citep{iashin2021taming}), FAD$_{edit}$ between the distribution of source and edited samples and CLAP$_{edit}$ between the edited audio and the editing prompt.
Subjective evaluations relate to (i) overall quality (OVL), and (ii) relevance to the text input
(REL), both using a Likert scale (from 1 to 5).
Additionally for music editing evaluations we report (iii) editing consistency (CON).
% where raters are asked to score the level of consistency between the editing and original samples, considering the provided editing prompt.
% how good of an edit is the edited sample of the original, considering the provided editing prompt.
Raters were recruited using the Amazon Mechanical Turk platform and all samples were normalized to -14dB LUFS \citep{series2011algorithms}.
For stereo samples objective evaluation the signal is down mixed into mono prior to metrics computation.
For subjective ratings we keep the original audio format generated by each model.
A screenshot of the evaluation form is presented in appendix \ref{apx:amt_evals}.
% For text-to-music evaluation we keep the original audio format generated by each model. For music editing evaluation we normalize all samples into 32 kHz mono format.
% We evaluate randomly sampled files, where each sample was evaluated by at least 5 raters.
% We use the CrowdMOS package7 to filter noisy annotations and outliers. We remove annotators who
% did not listen to the full recordings, annotators who rate the reference recordings less than 85, and the
% rest of the recommended recipes from CrowdMOS [Ribeiro et al., 2011]. For fairness, all samples
% are normalized at −14dB LUFS [ITU-R, 2017].

% \subsubsection{Quantitative metrics}
% \paragraph{FD}
% \paragraph{KL}
% \paragraph{CLAP}
% \subsubsection{Qualitative metrics}
% \paragraph{Audio quality}
% \paragraph{Text adherence}
% \paragraph{Edit fidelity}
% \paragraph{Musicality}
% \paragraph{Stereo correctness}
% \paragraph{Musical structure}
% \subsubsection{}

\section{Results}

% We report the model performance on text-guided music editing.
%Additionally we provide samples at \url{https://melodyflow.github.io/}.\wn{1. can we link this at the end of introduction or abstract? that's the first thing reviewers would look for}\wn{2.Also the title in the demo page is not updated (missing "Generation")}\wn{3. In demo page, let's highlight which column is the proposed method? It's hard to tell which one we advocates for (I assume the tedit=0.0625, but viewers won't know before reading the full paper)}\wn{4. can we also include generation samples (in addition to just editing samples)}
% \gael{@Zhaoheng/Varun to add link.}
% Ablations are conducted on the choice of latent representation, the impact of moving from mono to stereo, the transformer architecture and the model scale. We finally report memorisation results.
% For text-to-music qualitative evaluations we compare \proposed to three baselines: \musicgen, \audioldm2 and \stableaudio.
% For \musicgen, \audioldm2 we use the available open source implementations and for \stableaudio we use the public API (as of Wed. May 14 2024, AudioSpark 2.0 model version).
%\wn{include audioldm2 w/ ddpm samples on the demo page?}

\subsection{Text-guided music editing}

\begin{table}[t!]
  \centering\small
  % \vspace{-0.2cm}
  \caption{Comparison to baselines on text-guided high fidelity music editing of samples from the \textsc{in-domain} test set, using LLM-assisted editing prompts.}
  \label{tab:main_m2m_mmi}
  % \resizebox{0.9\columnwidth}{!}{
  \begin{tabular}{l|l|ccc|ccc|c}
    \toprule    
    \textsc{Model} &\textsc{Method}  & \textsc{Ovl.} $\uparrow$ & \textsc{Rel.} $\uparrow$ & \textsc{Con.} $\uparrow$& \textsc{Avg.} $\uparrow$\\
    \midrule
    % Reference        -      & -  & - &  & &  & -   \\
    % \midrule    
    % \audioldm 1    & DDPM inv. &  2.94  & 3.02 & 2.99  \\
    \audioldm2-music    & DDPM inv. &  2.48$\pm$0.07  & 2.36$\pm$0.08 & \textbf{2.72}$\pm$0.09 & 2.52 \\
    \musicgen-melody & Chroma cond. &  2.57$\pm$0.08 & 2.46$\pm$0.09 & 2.14$\pm$0.07 & 2.39 \\
    % \stableaudio & & ?  &  &   &   & &  \\
    \midrule
    % \proposed-medium & DDIM inv.  & 3.73 & 4.19 & 3.77 \\
    \proposed-medium & Reg. inv. & \textbf{2.72}$\pm$0.08 & \textbf{2.72}$\pm$0.07 & 2.61$\pm$0.10 & \textbf{2.68}\\
    \bottomrule
  \end{tabular}% }
  % \vspace{-0.5cm}
\end{table}

We compare \proposed-medium with existing open source music editing implementations, namely \musicgen-melody 
% (specifically trained with chroma conditioning for the task of text-conditioned music editing)
and \audioldm2 with DDPM inversion (following \citet{manor2024zero}).
% Although the \stableaudio API offers a music editing setting, for practical reasons we were not able to submit hundreds of samples hence exclude it from our baselines.
The Table \ref{tab:main_m2m_mmi} presents the main music editing subjective evaluation results.
% We compare \proposed-medium with our inversion method to \audioldm2 with DDPM inversion and \musicgen-melody.
\proposed outperforms both baselines on the quality and text-fidelity axes.
\musicgen-melody specifically underperforms consistency-wise while \audioldm2 suffers from lower text adherence.
Indeed during our listening tests we observe that \audioldm2 with DDPM inversion sometimes only generates a distorted version of the original track, hence does not take into account the editing prompt and keeps a strong similarity with the original.
This also explains why consistency-wise \proposed lags slightly behind \audioldm2.
Averaging on the three axes \proposed sets a new baseline for zero-shot music editing at test-time.

\subsection{Text-to-music generation}
\bsc{We can add some spectrograms of stereo vs. mono since this is one of our main contributions.}
Text-to-music generation performance is reported in the Table \ref{tab:main_t2m}.
For text-to-music qualitative evaluations we compare \proposed to three baselines that also support both generation and editing: \musicgen, \audioldm2, \stableaudio.
For \musicgen and \audioldm2 we use the available open source implementations and for \stableaudio we use the public API (as of Wed. May 14 2024, AudioSpark 2.0 model version).
\proposed achieves comparable performance with \musicgen, both lagging slightly behind \stableaudio in terms of human preference.
We do not report objective metrics on \stableaudio as none were reported on the full MusicCaps benchmark \cite{evans2024fast}.
We do not run any subjective evaluation against \magnet but report their objective metrics and latency values.
\proposed achieves remarkable efficiency with only 64 inference steps.
% to generate a 30 seconds sample.
% without any just in time compiling optimization.
% \stableaudio latency is reported from \citep{evans2024longform} although we observed an average latency of 33 seconds when querying the model through API.
% There may be additional components involved in the text-to-music generation process (such as text prompt refinement and post-processing that may improve overall performance compared with raw model outputs).
% Other latency measurements are reported from \citep{ziv2023masked}.

% \musicgen music_audioset_epoch_15_esc_90.14
% MusicFlow music_speech_epoch_15_esc_89.25.pt
% \audioldm2 music_speech_audioset_epoch_15_esc_89.98.pt
% \stableaudio 630k-audioset-fusion-best.pt

\begin{table}[t!]
  \scriptsize
  \centering
  % \vspace{-0.2cm}
  \caption{Comparison to text-to-music baselines. We report the original objective metrics for \audioldm2 and \musicgen. For subjective evaluations we report mean and CI95.}
  \label{tab:main_t2m}
  \begin{tabular}{lccc|cc|rrr}
    \toprule    
    \textsc{Model}       &FAD$_{\text{vgg}} \downarrow$      & \textsc{Kl} $\downarrow$ & CLAP$_{\text{sim}} \uparrow$ & \textsc{Ovl.} $\uparrow$ & \textsc{Rel.} $\uparrow$ &\textsc{\# Steps} & \textsc{Latency (s)}\\
    \midrule
    Reference           & -      & -  & - & 3.67$\pm$0.10 & 4.04$\pm$0.10 & -  & - \\
    \midrule    
    % Mousai               & 7.5       & \res{1.59}  & \res{0.23} &  &  & 200 & 44.0 \\
    % MusicLM               & 4.0       & -     & -  &   &  & - & -\\
    \audioldm 2     & 3.1  & \res{1.20}  & \res{0.31} & 2.79$\pm$0.08 & 3.40$\pm$0.08 & 208 & 18.1 \\
    \musicgen-small & 3.1  & \res{1.29}  & \res{0.31} & - & - & 1500 & 17.6\\  % 3.36$\pm$0.07 & 3.65$\pm$0.07
    \musicgen-medium  & 3.4  & \res{1.23}  & \res{0.32} & 3.40$\pm$0.08 & 3.79$\pm$0.07 & 1500 & 41.3\\
    \stableaudio &  -  & - &  - & 3.67$\pm$0.08 & 3.89$\pm$0.07 & 100 & 8.0 \\  % 8 seconds according to paper, 33 seconds measured between clic and getting output on their website. There may be preprocessing and postprocessing features included. They may use a slow GPU. \stableaudio AudioSparx 2.0 model.
    \magnet-small & 3.3  & \res{1.12}  & \res{0.31} & - & - & 180 & 4.0\\  % 3.36$\pm$0.07 & 3.65$\pm$0.07
    \magnet-large & 4.0  & \res{1.15}  & \res{0.29} & - & - & 180 & 12.6\\  % 3.36$\pm$0.07 & 3.65$\pm$0.07
    \midrule
    \proposed-small & 2.8   & 1.27 & 0.33 & 3.27$\pm$0.08 & 3.83$\pm$0.08 & 64 & 1.8 \\  % 3.7 for 10s samples on A100 (eager, no optimisation)
    \proposed-medium & 3.5   & 1.30  & 0.31 & 3.41$\pm$0.08 & 3.77$\pm$0.07 & 64 & 2.3 \\  % 9.2 for 30s samples on A100 (eager, no optimisation), 7.7 on H100.
    \bottomrule
  \end{tabular}
  % \vspace{-0.5cm}
\end{table}

\subsection{Latent Inversion}
\subsubsection{Inversion methods}
\label{exp:inv_methods}

\begin{figure}[t!]
    % \vspace{-0.2cm}
     \centering
     % \begin{subfigure}[b]{0.32\textwidth}
     %     \centering
     %     \includegraphics[width=\textwidth]{figs/cfg.pdf}
     %     \caption{CFG ablation.}
     %     \label{fig:thr_bz}
     % \end{subfigure}
     % \hfill
     \begin{subfigure}[b]{0.32\textwidth}
         \centering
         \includegraphics[width=\textwidth]{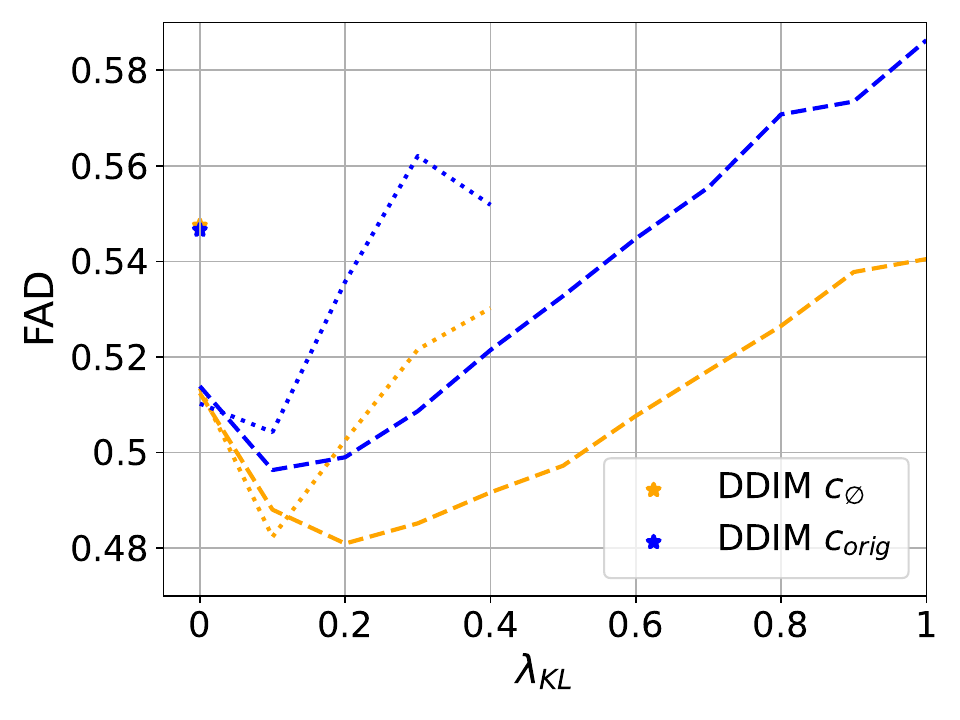}
         \caption{FAD$_{edit}$ as a function of $\lambda_{KL}$.}
         \label{fig:lkl_fad}
     \end{subfigure}
     % \hfill
     \begin{subfigure}[b]{0.32\textwidth}
         \centering
         \includegraphics[width=\textwidth]{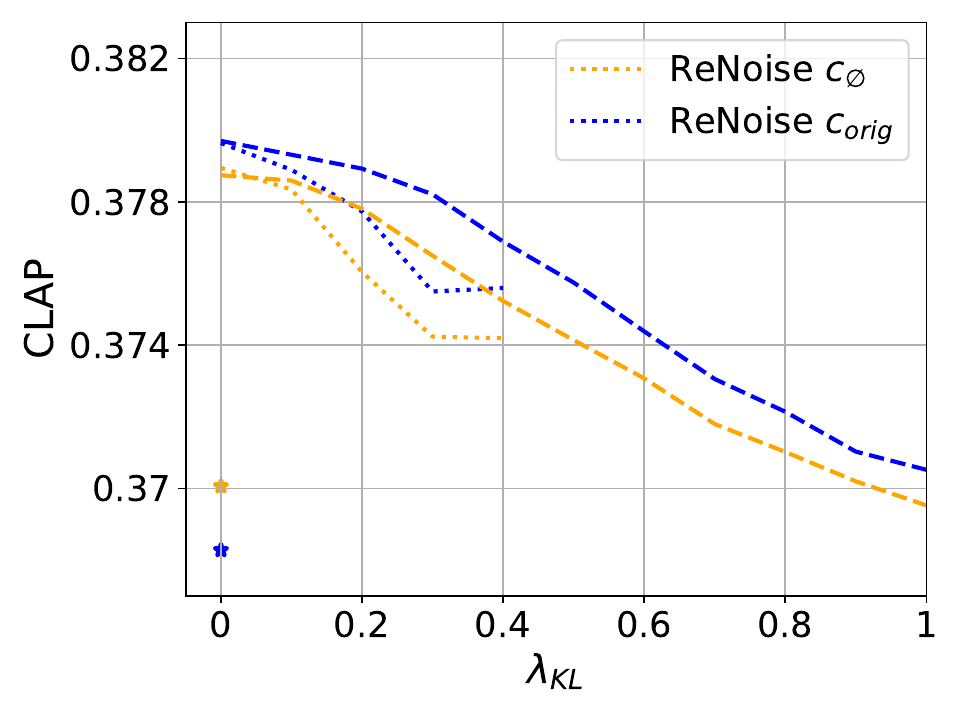}
         \caption{CLAP$_{edit}$ as a function of $\lambda_{KL}$.}
         \label{fig:lkl_clap}
     \end{subfigure}
     % \hfill
     \begin{subfigure}[b]{0.32\textwidth}
         \centering
         \includegraphics[width=\textwidth]{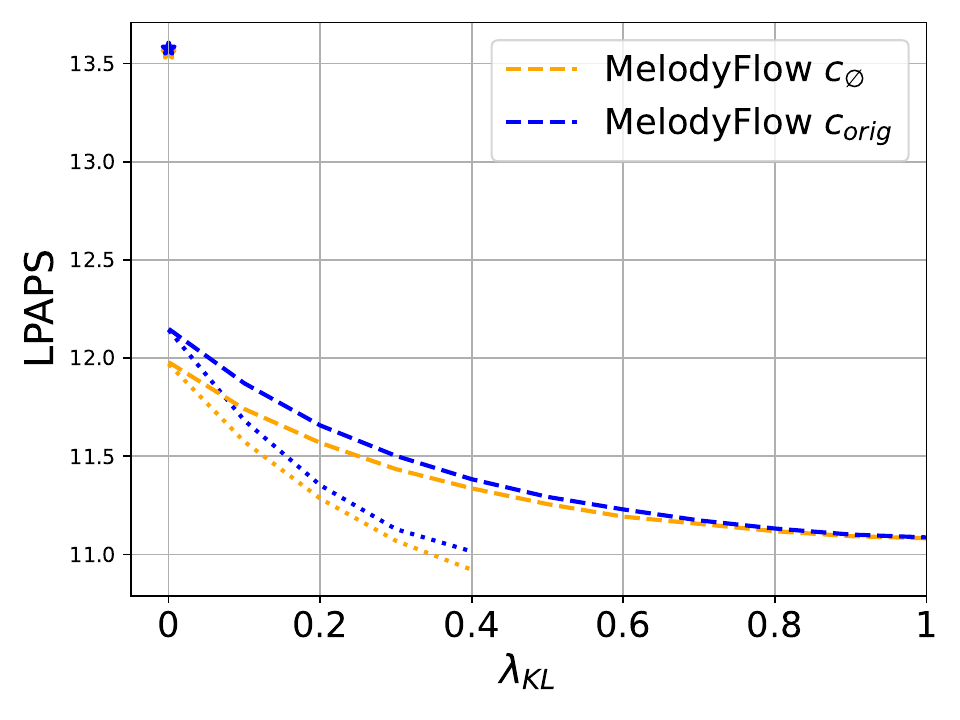}
         \caption{LPAPS as a function of $\lambda_{KL}$.}
         \label{fig:lkl_mse}
     \end{subfigure}
     % \hfill
     \caption{Effect of the regularization weight $\lambda_{KL}$ on the quality (Figure \ref{fig:lkl_fad}) and text-adherence (Figure \ref{fig:lkl_clap}) of music editing. $\epsilon$- and $v$-prediction are compared with or without $c_{orig}$.}
     % \vspace{-0.2cm}
\end{figure}

We compare \proposed with DDIM and \renoise in the Figures \ref{fig:lkl_fad}, \ref{fig:lkl_clap} and \ref{fig:lkl_mse}, as a function of the divergence loss weight $\lambda_{KL}$.
During the inversion we use a classifier-free-guidance (CFG) of 0 and employ a CFG of 4 during the regeneration.
The choice of zero CFG is meant to prevent divergence during inversion (see \ref{apx:cfg}).\bsc{Ablation on guidance weight?}\gl{in appendix} For \renoise and \proposed the predictions are regularized by the weighted KL patch-wise divergence loss $\mathcal{L}_{patchKL}$ of Algorithm \ref{alg:cap} and \renoise additionally uses an autocorrelation loss with $\lambda_{pair} = 10$ \citep{garibi2024renoise}.
Both also employ the reversible inversion trajectory estimation while DDIM does not.
% In the Figures \ref{fig:lkl_fad} and \ref{fig:lkl_clap} we ablate on the divergence loss weight $\lambda_{KL}$ for $T_{edit}=0.04$, using $S=25, K=4, w_k=k-1$.
% Since during latent inversion we know the original latent $\mathbf{x}_{orig}$, the model prediction can be rewritten as $\epsilon$-prediction by computing $\epsilon = v_{\Theta}(\mathbf{z}_t,t,c) - \mathbf{x}_{orig}$.
% In that scenario we can apply the exact ReNoise inversion method of \citep{garibi2024renoise}, using 10 iterations of noise de-correlation (shown as $\epsilon$-prediction in the Figures).
% Whether the model prediction is expressed as noise or velocity, 
The Figures show that both \proposed and \renoise outperform DDIM inversion by a large margin on the three evaluated axes.
an optimum can be achieved around $\lambda_{KL} = 0.2$ for velocity prediction and around $0.1$ for noise prediction.
Overall the quality is better (lower FAD$_{edit}$ in the Figure \ref{fig:lkl_fad}) when directly regularizing the velocity prediction.
In both cases we observe a higher CLAP$_{edit}$ in the Figure \ref{fig:lkl_clap} when the original text description $c_{orig}$ conditions the inversion process, confirming better text-adherence.
This happens at the expense of a higher FAD$_{edit}$ compared with unconditional inversion.

\subsubsection{Target inversion flow step}
\label{exp:inv_target}
\begin{figure}[t!]
    % \vspace{-0.2cm}
     \centering
     % \begin{subfigure}[b]{0.32\textwidth}
     %     \centering
     %     \includegraphics[width=\textwidth]{figs/cfg.pdf}
     %     \caption{CFG ablation.}
     %     \label{fig:thr_bz}
     % \end{subfigure}
     % \hfill
     \begin{subfigure}[b]{0.32\textwidth}
         \centering
         \includegraphics[width=\textwidth]{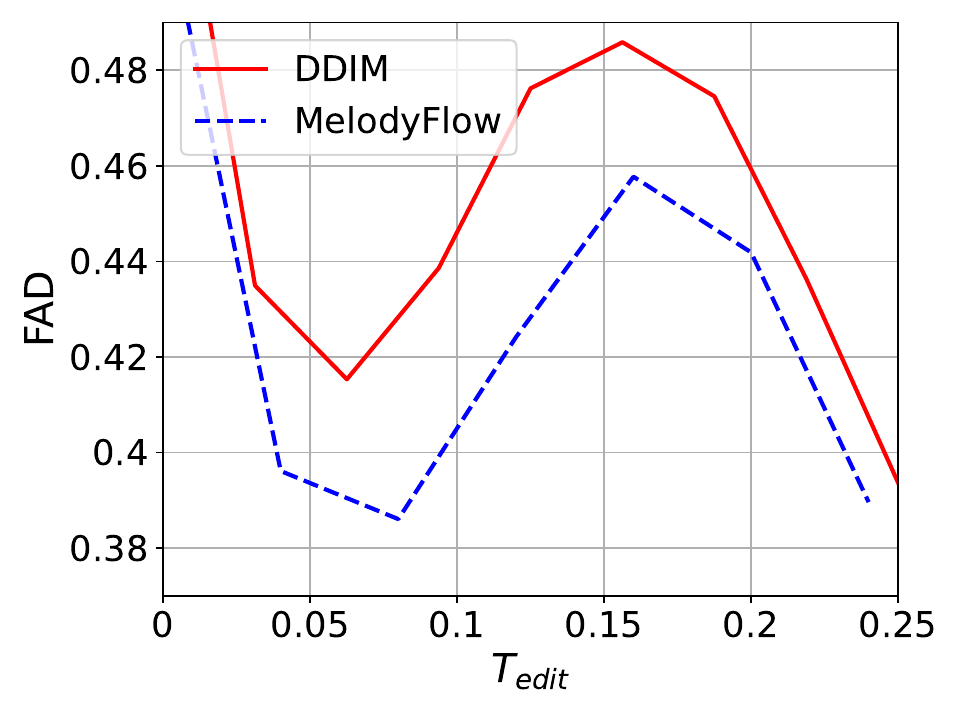}
         \caption{Editing FAD$_{edit}$ vs. $T_{edit}$.}
         \label{fig:tgt_fad}
     \end{subfigure}
     % \hfill
     \begin{subfigure}[b]{0.32\textwidth}
         \centering
         \includegraphics[width=\textwidth]{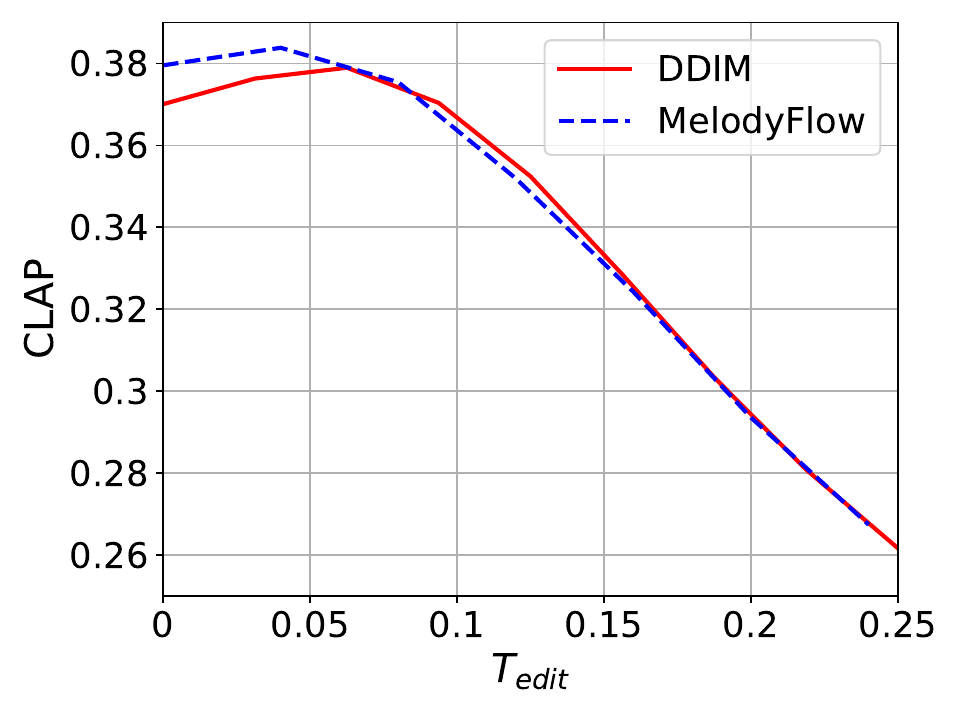}
         \caption{Editing CLAP$_{edit}$ vs. $T_{edit}$.}
         \label{fig:tgt_clap}
     \end{subfigure}
     % \hfill
     \begin{subfigure}[b]{0.32\textwidth}
         \centering
         \includegraphics[width=\textwidth]{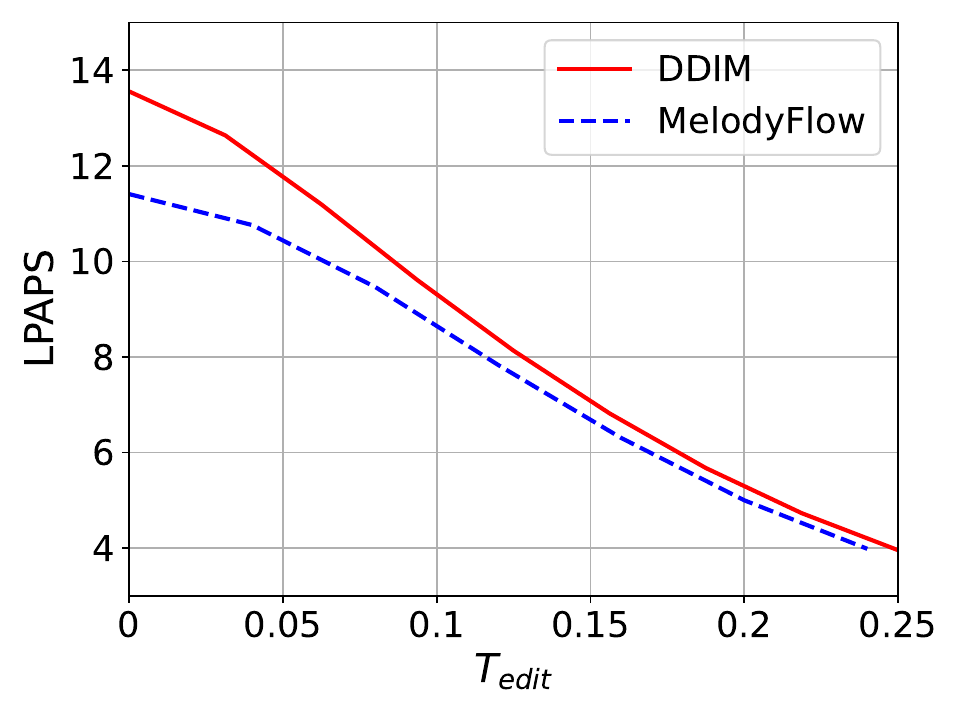}
         \caption{Editing LPAPS vs. $T_{edit}$.}
         \label{fig:tgt_mse}
     \end{subfigure}
     % \vspace{-0.2cm}
     \caption{Music editing quality as a function of the target inversion step $T_{edit}$. We report FAD$_{edit}$ (Figure \ref{fig:tgt_fad}), CLAP$_{edit}$ (Figure \ref{fig:tgt_clap}) and LPAPS (Figure \ref{fig:tgt_mse}) objective metrics.}% act as a proxy for quality, text-adherence and consistency, respectively.}
     %Efficiency quality trade offs of \proposed in the text-guided music generation and editing settings. The combination of our flow matching design choices enable faster generation for a given efficiency budget or better overall quality. Editing quality objective metrics indicate a sweet spot around 128 NFE.}
     % \vspace{-0.2cm}
\end{figure}

%In this work we focus on test time and fine tuning free editing methods. The most straightfor DDIM and ReNoise.
In the Figures \ref{fig:tgt_fad}, \ref{fig:tgt_clap} and \ref{fig:tgt_mse} we report music editing objective metrics as a function of $T_{edit}$, comparing DDIM inversion with \proposed.
The consistency with the source sample is higher (lower LPAPS) with our method than DDIM inversion.
% , which correlates with our listening experience.
The S-shaped FAD curves of the Figure \ref{fig:tgt_fad} indicate an inversion optimum around $T_{edit}=0.06$, correlating with the peak in CLAP$_{edit}$ score.

\bsc{Section~\ref{exp:codec} and "Model training" subsection should be put later than editing ablation }\gl{done}
\subsection{Codec bottleneck regularizer}
\label{exp:bottleneck}

% To better understand the impact of codec bottleneck and frame rate on the generative model performance,
All other things being equal, we ablate on the bottleneck regularizer for a fixed frame rate of 50 Hz by comparing RVQ- (using 4 codebooks of size 2048 each), KL-regularizer \citep{evans2024fast} and no regularizer at all in the Table~\ref{tab:abl_design_codec}.
Our results indicate optimal reconstruction performance with no regularizer, closely followed by KL. RVQ stands much further away, likely due to the high level of compression enforced by the discretization (despite the significant dictionary size of $2048^4 = 1.7\times10^{13}$).
The same ranking applies for SI-SDR \citep{le2019sdr}.
Regarding text-to-music generation performance, the KL-regularizer outperforms the other options.
Overall this shows the KL-regularizer offers the best trade off between reconstruction and generation performance.
% Regarding FM model performance we report a decreasing in-domain FAD in the favor the VAE.
% , that is confirmed by human evaluations.

Ablating on the codec frame rate with the KL regularizer shows that 5 Hz achieves comparable performance with the 50 Hz codecs trained with RVQ or no regularizer, a 10$\times$ improvement factor.
% FM model performance is closer to that of 50 Hz codec with identity bottleneck.
We chose to work with the 20 Hz KL-regularized codec for the 32 kHz mono \proposed-small, as it offers a good trade off between quality and speed.
Accounting for the additional information to compress when scaling to 48 kHz stereo, we chose a frame rate of 25 Hz for \proposed-medium.

% as a good compromise between quality and generative model efficiency.
% Finally we ablate on the VAE codec frame rate by experimenting with 20 and 5 Hz instead.
% At 5 Hz we report similar compression performance to the 50 Hz codec with RVQ, showing that the frame rate can only be decreased up to a limit between 5 and 20 Hz before affecting the generative model performance.

% Given that removing the quantizer boosts the reconstruction performance compared with RVQ codec versions, we investigate the impact of codec framerate and VAE on the text-to-music task. We compare \proposed with the following codec versions: RVQ mono at 50Hz (\musicgen setting, where latents are extracted before quantization is applied), open-source encodec (MusicFlow setting), quantizer-free mono at 50Hz, VAE mono at 50Hz and VAE mono at 10Hz (where the reconstruction performance is similar to RVQ 50Hz).

\begin{table}[t!]
  \centering\small
  \caption{\label{tab:abl_design_codec}Codec bottleneck and framerate ablation for 32 kHz mono audio. Both compression and generative model performances are reported on the \textsc{in-domain} test set.}
  % \resizebox{0.9\columnwidth}{!}{
  %\setlength{\tabcolsep}{3pt}
  \begin{tabular}{lc|cr|ccc}
    \toprule    
    \textsc{Regularizer} &\textsc{Frame rate (Hz)} &\textsc{STFT}$_{loss} \downarrow$ &\textsc{SI-SDR}$\uparrow$ &FAD$_{vgg} \downarrow$ \\
    \midrule
    \textsc{$\varnothing$} & 50 & 0.35 & 18.5 & 0.68 \\
    \midrule
    \textsc{RVQ} & 50 & 0.55 & 4.4 & 0.55 \\
    \midrule
     & 50 & 0.34 & 18.1 & 0.48 \\
    \textsc{KL} & 20 & 0.44 & 12.9 & 0.47 \\
     & 5 & 0.53 & 3.5 & 0.67 \\
    \bottomrule
  \end{tabular}%}
  % \vspace{-0.4cm}
\end{table}

\subsection{FM design}

\subsubsection{Model training}
\label{exp:design}

We compare our FM model design with the baseline implementation of \citet{le2024voicebox}, both being trained on the same music latents.
The most notable changes are the removal of the infilling objective during training, the change in flow step sampling and the introduction of mini-batch coupling.
Table \ref{tab:abl_design} presents the impact of those choices on the last FM model validation \textsc{MSE$_{loss}$} of the EMA checkpoint, and on the in-domain test FAD (in the text-to-music generation setting).
No loss value is reported for the baseline as the infilling objective facilitates the task (hence values are not fairly comparable), and for validation we sample flow steps uniformly regardless of the training sampling scheme. 
Such infilling objective in \citet{le2024voicebox}'s FM model was designed to handle variable length sequences that are inherent to the speech domain.
In our experiments it showed to be detrimental for the model performance, and we know diffusion models can support infilling/outfilling without additional tweaks \citep{liu2023audioldm}.
With all methods combined the in-domain FAD is reduced to 0.39 from 0.53 and consistent with the observed loss decrease, which validates our design.

\begin{table}[t!]
  \centering\small
  \caption{FM model design ablation. FAD (resp. MSE) is reported on the \textsc{in-domain} test (resp. validation) set. Baseline is adapted from \citep{le2024voicebox} but retrained on our music latents.}
  \label{tab:abl_design}
  % \resizebox{0.9\columnwidth}{!}{
  % \setlength{\tabcolsep}{3pt}
  \begin{tabular}{lccccc|cc}
    \toprule    
    \textsc{Ablation}  & \textsc{Heads} & \textsc{Layers} & \textsc{Infill} & \textsc{Sampling} & \textsc{OT-FM} & \textsc{MSE}$_{loss}$ $\downarrow$ & FAD$_{vgg}$    $\downarrow$ \\
    \midrule
    baseline & 16 & 24 & \ding{51} & uniform & \ding{55} & - & .53   \\
    $-$ infilling & 16 & 24 & \ding{55} & uniform & \ding{55}  & .8596 & .50 \\
    $+$ sampling & 16 & 24 & \ding{55} & logit-normal & \ding{55}  & .8484 & .44 \\
    % $+$ DiffiT cond. & .45 & .8460 \\
    % $+$ Variance rescaling & .47 & .8462  \\
    $+$ batch coupling & 16 & 24 & \ding{55} & logit-normal & \ding{51}  & .8322 & .42 \\
    $+$ wider model & 18 & 18  & \ding{55} & logit-normal & \ding{51}  &.8310 & .39 \\
    \bottomrule
  \end{tabular}% }
  % \vspace{-0.4cm}
\end{table}

\begin{figure}[t!]
    % \vspace{-0.2cm}
     \centering
     % \begin{subfigure}[b]{0.32\textwidth}
     %     \centering
     %     \includegraphics[width=\textwidth]{figs/cfg.pdf}
     %     \caption{CFG ablation.}
     %     \label{fig:thr_bz}
     % \end{subfigure}
     % \hfill
     \begin{subfigure}[b]{0.32\textwidth}
         \centering
         \includegraphics[width=\textwidth]{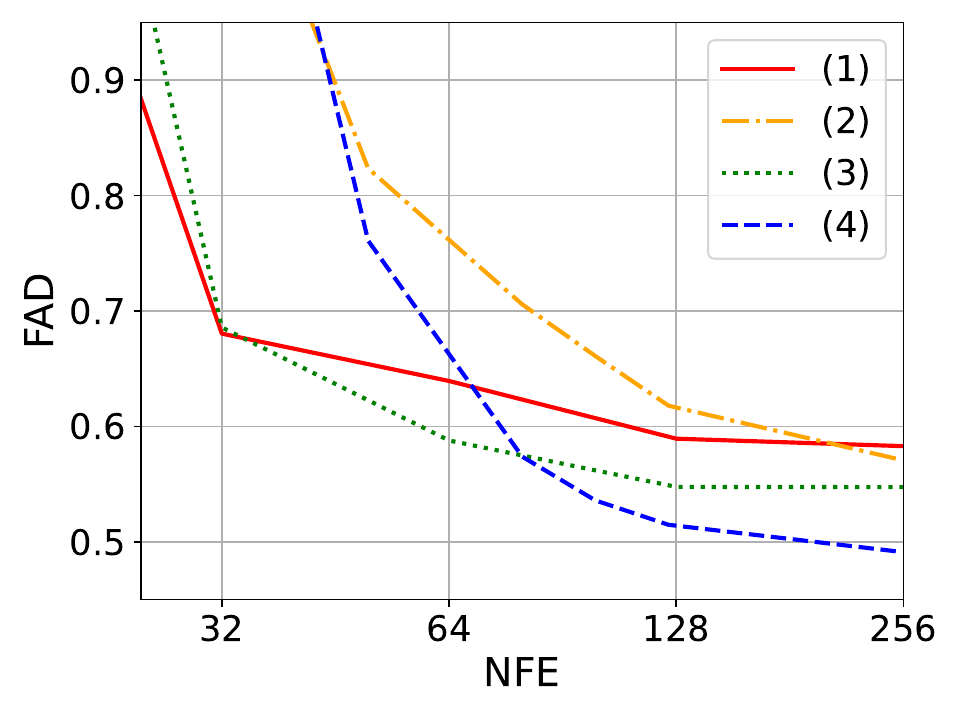}
         \caption{FAD$_{edit}$ vs. NFE.}
         \label{fig:fad_nfe}
     \end{subfigure}
     % \hfill
     \begin{subfigure}[b]{0.32\textwidth}
         \centering
         \includegraphics[width=\textwidth]{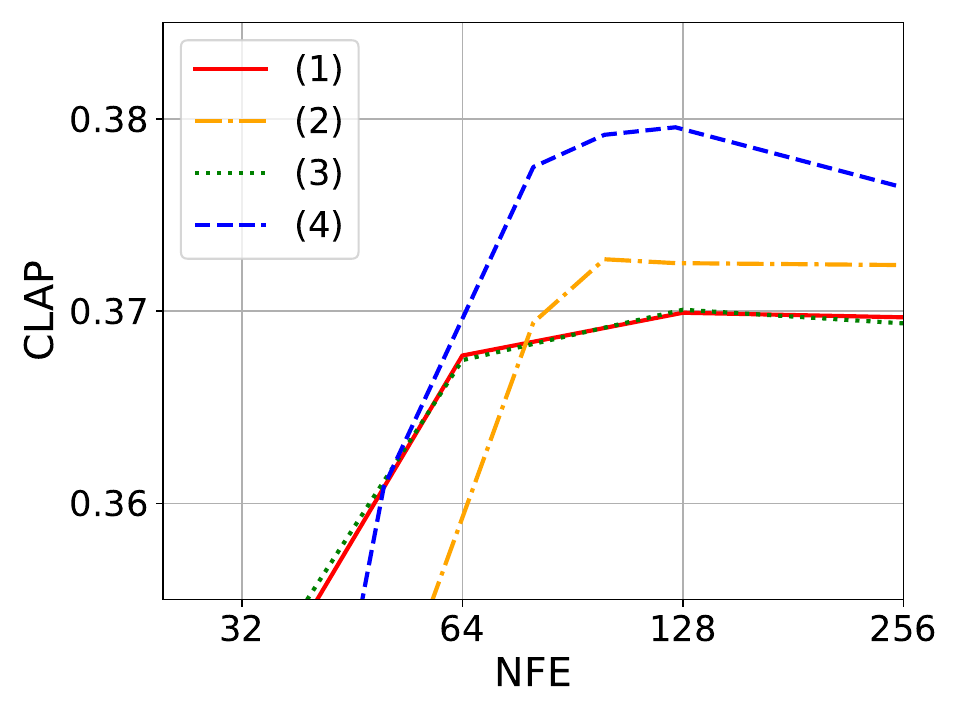}
         \caption{CLAP$_{edit}$ vs. NFE.}
         \label{fig:clap_nfe}
     \end{subfigure}
     % \hfill
     \begin{subfigure}[b]{0.32\textwidth}
         \centering
         \includegraphics[width=\textwidth]{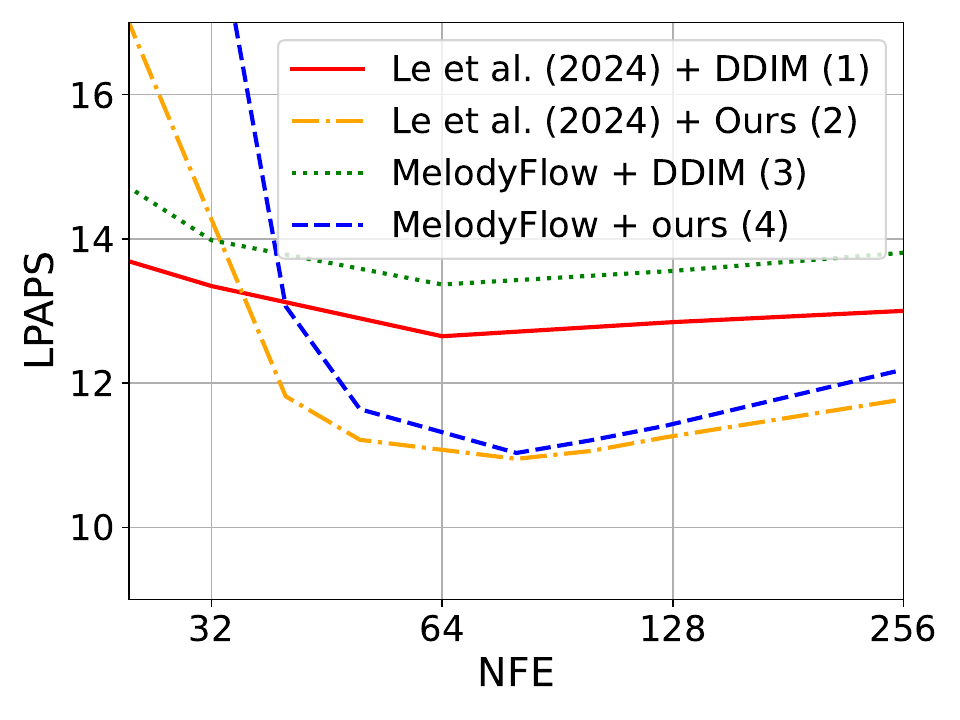}
         \caption{LPAPS vs. NFE.}
         \label{fig:lpaps_nfe}
     \end{subfigure}
     % \vspace{-0.2cm}
     \caption{Efficiency-quality trade offs of \proposed in the text-guided music editing setting, measured using objective metrics. Objective metrics (FAD$_{edit}$ in the Figure \ref{fig:fad_nfe}, CLAP$_{edit}$ in the Figure \ref{fig:clap_nfe} and LPAPS in the Figure \ref{fig:lpaps_nfe}) indicate a sweet spot around 128 NFE.\bsc{1-4 are not annotated}\gl{done}}
     % \vspace{-0.2cm}
\end{figure}

\subsubsection{Music editing quality/efficiency}
\label{exp:inversion}
%In this work we focus on test time and fine tuning free editing methods. The most straightfor DDIM and ReNoise.
We compare DDIM with \proposed's inversion using a target inversion step of $T_{edit}=0$. FAD$_{edit}$ (Figure \ref{fig:fad_nfe}), CLAP$_{edit}$ (Figure \ref{fig:clap_nfe}) and LPAPS (Figure \ref{fig:lpaps_nfe}) are plotted as a function of the total NFE count (inversion + regeneration included).
%A good trade off between speed, quality, text-adherence and consistency is achieved at 128 NFEs.
Quality-wise, the combination of our FM and inversion designs outperform the baseline.
Regardless of the FM design choice, DDIM inversion requires as few as 32 NFEs to achieve an acceptable FAD.
Our inversion only outperforms after 125 NFEs.
On the text-adherence axis, the FM model design alone does not translate in better performance when combined with DDIM inversion.
Swapping DDIM with our method shows a different trend, highlighting the benefit of combining FM and inversion methods.
Analyzing the consistency with the original sample, again we observe that the regularized inversion plays a more important role than the FM model design: the baseline FM model actually outperforms ours when used in conjunction with DDIM inversion.
Overall our method consistently outperform the baseline for 125 NFEs.

\section{Related work}
\subsection{Audio representation}

% Most recent works rely on compressing the waveform into a sequence of latent representations (either discrete or continuous), on which a generative model is trained.
% One of the first discrete representations were k-means indexes extracted from high level features of a transformer encoder \citep{hsu2021hubert}, for speech language modeling.

Recent advancements in neural codecs have seen the application of VQ-VAE on raw waveforms, incorporating a RVQ bottleneck as demonstrated in \cite{zeghidour2021soundstream,defossez2022high}, later refined as per \cite{kumar2024high}. \citet{evans2024fast} proposed a modification to this approach by replacing the RVQ with a VAE bottleneck to enhance the modeling of continuous representations.
In addition, several recent audio generative models have adopted Mel-Spectrogram latent representations, coupled with a vocoder for reconstruction, as shown in the works of \citep{ghosal2023text,liu2023audioldm2,le2024voicebox}.
%  Furthermore, \cite{defossez2022high} introduced an additional layer of complexity by incorporating quantization to support discrete representation on top of the continuous representation.

\subsection{Text-to-music generation}

% Several generative model architectures were proposed for the task of text-to-music generation, relying on either discrete or continuous audio representations.
Models that operate on discrete representation are presented in the works of \citep{agostinelli2023musiclm,copet2024simple,ziv2023masked}. \cite{agostinelli2023musiclm} proposed a representation of music using multiple streams of tokens, which are modeled by a cascade of transformer decoders conditioned on a joint textual-music representation \citep{huang2022mulan}.
\cite{copet2024simple} introduced a single-stage language model that operates on streams of discrete audio representations, supporting both 32 kHz mono and stereo. \cite{ziv2023masked} replaced the language model with a masked generative single-stage non-autoregressive transformer.
\cite{schneider2023mo,huang2023noise2music,liu2023audioldm2} use diffusion models. \cite{schneider2023mo} utilized diffusion for both the generation model and the audio representation auto-encoder. \cite{liu2023audioldm2} trained a foundational audio generation model that supports music with latent diffusion, conditioned on autoregressively generated AudioMAE features \citep{huang2022masked}.
\cite{evans2024fast,evans2024longform} proposed an efficient long-form stereo audio generation model based on the latent diffusion of VAE latent representations. This model introduced timing embeddings conditioning to better control the content and length of the generated music.

\subsection{Music editing}

% \gael{Distinguish between inpainting, controls, instructions, and style transfer?}

% \cite{wang2023audit} train an instruction-guided audio editing model based on latent diffusion models. The model only requires edit instructions as opposed to full audio descriptions of the expected output audio. 

\cite{lin2024arrange} proposed a parameter-efficient fine-tuning method for autoregressive language models to support music inpainting tasks.
% such as track-conditioned music refinement and score-conditioned music arrangement. Their model achieves comparable quality to unconditioned autoregressive generation.
\cite{garcia2023vampnet} developed a masked acoustic modeling approach for music inpainting, outpainting, continuation and vamping.
\cite{wu2023music} fine-tuned a diffusion-based music generation model with melody, dynamics and rhythm conditioning.
% introduced a diffusion-based music generation model that offers multiple precise, time-varying controls over generated audio. They specifically fine tune a conditional model on melody, dynamics and rhythm controls.
\cite{novack2024ditto} is a fine-tuning free framework for controlling pre-trained text-to-music diffusion models at inference-time via initial noise latent optimization. %  It supports inpainting, outpainting, looping as well as intensity, melody, and musical structure control.
\cite{zhang2024musicmagus} investigated zero-shot text-guided music editing with conditional latent space and cross attention maps manipulation.
% In relies on word swapping in the original text description of the input audio along with conditional signal space and cross attention maps manipulation.
\cite{manor2024zero} employs DDPM inversion \citep{huberman2023edit} for zero-shot unsupervised and text-guided audio editing. %  with \audioldm2.

% \gael{DITTO: Diffusion Inference-Time T -Optimization for Music Generation.}

% \cite{zhang2024musicmagus} focuses on zero-shot text-guided music editing. In relies on word swapping in the original text description of the input audio along with conditional signal space and cross attention maps manipulation.
% a Large Language Model to generate a large number of paired captions of a specific target editing task to estimate the corresponding conditioning signal bias.

% \cite{manor2024zero} employs DDPM inversion to perform zero-shot unsupervised and text-guided audio editing with AudioLDM2.

% AUDIT: Audio Editing by Following Instructions with Latent Diffusion Models

% INVESTIGATING PERSONALIZATION METHODS IN TEXT TO MUSIC GENERATION
% InstructME: An Instruction Guided Music Edit And Remix Framework with Latent Diffusion Models

% \subsection{Metrics}

% \cite{paissan2023audio} \gael{LPAPS definition.}
% \cite{gui2024adapting} Adapting frechet audio distance for generative music evaluation.

% \subsubsection{Autoregressive models}
% \subsubsection{Non-autoregressive models}
% \subsubsection{End-to-end diffusion models}
% \subsubsection{Spectrogram diffusion models}
% \subsubsection{Latent diffusion models}
%\subsection{High sampling rate and stereo generation}
%\subsubsection{Text embeddings}
%\subsubsection{Fast generation of variable-length, long-form audio}
%\subsubsection{Timing conditioning}
%\subsection{Evaluation metrics}
%\subsubsection{Multitask generative modeling}

\section{Discussion}

\paragraph{Limitations}

The proposed model specifically focuses on text-guided audio editing with the quality/efficiency trade off in mind, hence we do to not aim nor claim to outperform previous state of the art text-to-music generation models.
Under our current setup text-guided music editing prompts are not instructions.
They describe what the edited sample should sound like given an original music sample and description, but the model is not designed to understand direct editing instructions like \emph{replace instrument A by instrument B}.
While we observed that \proposed performs convincing editing tasks for several axes (genre or instrument swap, tempo modification, key transposition, inpainting/outpainting), more research work is required to accurately evaluate each of those axes.
Music editing human listening tests are conducted for a fixed $T_{edit}$, but eventually it should depend on the sound designer's preference on the creativity-consistency axis.
%the choice of $T_{edit}$ and inversion algorithm should depend on the sound designer's preference on the creativity-consistency axis.
% This is also related to data preparation where applying better formatting (rather than just using raw descriptions from the training dataset) may help.
% Eventually this is a matter of whether such model is meant for professional sound designers.
Finally the reported objective metrics are mostly used as a proxy for subjective evaluations but they have their limitations.
As an example we observe that optimizing FAD for MusicCaps is usually achieved by overfitting on our training dataset, which negatively correlates with perceived quality.
% (this is notably observed for several models of bigger sizes compared with their smaller variant).
Overall subjective evaluations remain the best source of truth until a model that mimics human ratings is developed.
% and does not reflect the optimal subjective performance, which is better correlated with optimizing on our internal evaluation dataset.

% \paragraph{Broader Impact}
% \label{ref:impact}
% Such a model has the potential to help creators expand their creativity by generating different styles of high quality music.
% The combination of efficiency and quality may eventually revolutionise the music sound design and composition work.
% This should unlock the ability for any user to listen to the music they like under a different style.
% With the features offered by such a model any user would be able to listen to the music they like under a different style.
% However such a model can also give malicious users the ability to alter and republish any copyrighted content without the original creators content.

\paragraph{Conclusion}

In this work we presented \proposed, the first non-autoregressive model tailored for zero-shot test-time text-guided editing of high-fidelity stereo music.
In the text-to-music setting the model offers competitive performance thanks to a low frame rate VAE codec and FM model featuring logit-normal flow step sampling, optimal-transport minibatch coupling and L-shaped attention mask.
% By combining logit-normal flow step sampling, optimal-transport minibatch coupling and L-shaped attention mask we train a versatile single-stage flow matching that - 
Combined with our proposed regularized latent inversion method, \proposed outperforms previous zero-shot test-time methods by a large margin.
The model achieves remarkable efficiency that is key for the sound design creative process and supports variable duration samples.
Our extensive evaluation, that includes objective metrics and human studies, highlights \proposed promise for efficient music editing with remarkable consistency, text-adherence and minimal quality degradation compared with the original, while remaining competitive on the task of text-to-music generation.
For future work we intend to explore how to accurately evaluate specific editing axes and how such a model could help design metrics that better correlate with human preference.

%The model offers a good quality efficiency trade off thanks to a low frame rate VAE codec, combined with logit-normal flow step sampling, optimal-transport minibatch coupling and L-shaped attention mask in the flow matching model.

% \input{sections/7_appendix}

% \input{sections/7_appendix}
% \input{sections/5_related}



\bibliographystyle{iclr2025_conference}
\bibliography{neurips_2024}

\appendix
% \section{Appendix}
\section{Appendix}

% \subsection{\proposed key components}
% \label{apx:details}

%\gael{maybe add reference to audio editing paper that may share similar observations}

% \gael{Explain that DDIM inversion does not accurately reconstruct}.
% The editing algorithm is presented in \ref{alg:renoise}.

\subsection{Experimental setup}
% \subsubsection{Models and hyperparameters}
\label{apx:exp_setup}
\subsubsection{Audio latent representation}
\label{apx:latent}

Our compression model implementation is that of \cite{copet2024simple}\footnote{\url{https://github.com/facebookresearch/audiocraft/blob/main/audiocraft/models/encodec.py}} enhanced by band-wise discriminators and snake activations from \citet{kumar2024high}, perceptual weighting \citep{wright2019perceptual}, VAE bottleneck and multi resolution STFT reconstruction loss from \citet{evans2024fast}.
We train a mono 32 kHz codec at 20 Hz frame rate and another one supporting stereo 48 kHz audio at 25 Hz. The bottleneck dimension is of 128. Both are trained on one-second random audio crops for 200K steps, with a constant learning rate of 0.0003, AdamW optimizer and loss balancer of \citep{defossez2022high}.
Stereo codecs are trained with sum and difference loss \citep{steinmetz2020automatic}.
The bottleneck layer statistics are tracked during training (dimension-wise) for normalization prior to FM model training.

% Our compression model iterates over that of \cite{copet2024simple} by adding a few upgrades. Namely we add bandwise discriminators and snake activations \citep{kumar2024high}, perceptual weighting \citep{wright2019perceptual}, VAE bottleneck \citep{evans2024fast} and use a multi resolution STFT reconstruction loss instead of MelSpectrogram.
% We experiment with a mono 32 kHz codec at 20 Hz frame rate and another one supporting stereo 48 kHz audio at 25 Hz.
% The bottleneck dimension is 128 and it is made of a VAE (the output dimension of the encoder is doubled to materialize the estimated mean and standard deviation).
% All codecs are trained on one-second random audio crops for 200K steps, with a constant learning rate of 0.0003, AdamW optimizer and loss balancer of \citep{defossez2022high}.
% Stereo codecs are trained with sum and difference loss \citep{steinmetz2020automatic}.

\subsubsection{Flow matching model}
\label{apx:fm}

\proposed's DiT follows \cite{esser2024scaling} configurations where each head dimension is of 64 and the model has the same number of heads and layers (either 18 or 24).
Model implementation is that of \verb+audiocraft+\footnote{\url{https://github.com/facebookresearch/audiocraft/blob/main/audiocraft/modules/transformer.py}} but adapted for FM following \citet{vyas2023audiobox}: U-shaped skip connections are added along with linear projections applied after concatenation with each transformer block output \cite{Bao_2023_CVPR}.
The model is conditioned via cross attention on a T5 representation \citep{JMLR:v21:20-074} computed from the text description of the audio, using 20\% dropout rate during training in anticipation for the classifier free guidance applied at inference.
Cross attention masking is used to properly adapt to the text conditioning sequence length of each sample within a batch and we use zero attention for the model to handle unconditional generation transparently.
No prepossessing is applied on the text data and we only rely on the descriptions (additional annotations tags such as musical key, tempo, type of instruments, etc. are discarded, although they also sometimes appear in the text description).
The flow timestep is injected following \citet{hatamizadeh2023diffit}. Minibatch coupling is computed with \verb+torch-linear-assignement+\footnote{\url{https://github.com/ivan-chai/torch-linear-assignment}}.
\proposed-small (resp. \proposed-medium) is trained on latent representation sequences of 32 kHz mono (resp. 48 kHz stereo) segments of 10 (resp. 30) seconds, encoded at 20 Hz frame rate (resp. 25 Hz).
% \proposed-small are trained on 48 kHz stereo segments of 30 seconds (encoded at 25Hz frame rate).
% For ablations we also consider codecs operating from 5 to 50 Hz frame rates.
\proposed-small (resp. \proposed-medium) is trained for 240k (resp. 120k) steps with AdamW optimizer ($\beta_1=0.9$, $\beta_2=0.95$, weight decay of 0.1 and gradient clipping at 0.2), a batch size of 576 and a cosine learning rate schedule with 4000 warmup steps.
Additionally, we update an exponential moving average of the model weights ever 10 steps with a decay of 0.99.
Each model is trained on 8 H100 96GB GPUs with \verb+bfloat16+ mixed precision and FSDP \citep{zhao2023pytorch}. \proposed-small requires 3 days and \proposed-medium 6 days of training.

\subsubsection{LLM-assisted editing prompt generation}
\label{ref:llm_assist}

For editing prompts design we prompted the LLama-3 large language model \cite{dubey2024llama} to modify the original descriptions by targeting genre swapping.
Edited descriptions were then manually verified to ensure their plausibility and coherence.
As an example, given the original description \emph{This is a lush indie-folk song featuring soaring harmony interplay and haunting reverb-y harmonica}, the resulting editing prompt is \emph{This is a lush Indian classical-inspired song featuring soaring harmony interplay and haunting reverb-y bansuri flute}.
% for the purpose of modifying the original descriptions in accordance with the specified target genre. 
% We did not solely rely on the automated system but we also manually verified the modifications to ensure their plausibility and coherence. As an example, given the original description "This is a lush indie-folk song featuring soaring harmony interplay and haunting reverb-y harmonica", the resulting editing prompt is "This is a lush Indian classical-inspired song featuring soaring harmony interplay and haunting reverb-y bansuri flute".

\subsubsection{Subjective evaluation form}
\label{apx:amt_evals}

\begin{figure}[t!]
    % \vspace{-0.2cm}
     \centering
     \includegraphics[width=\textwidth]{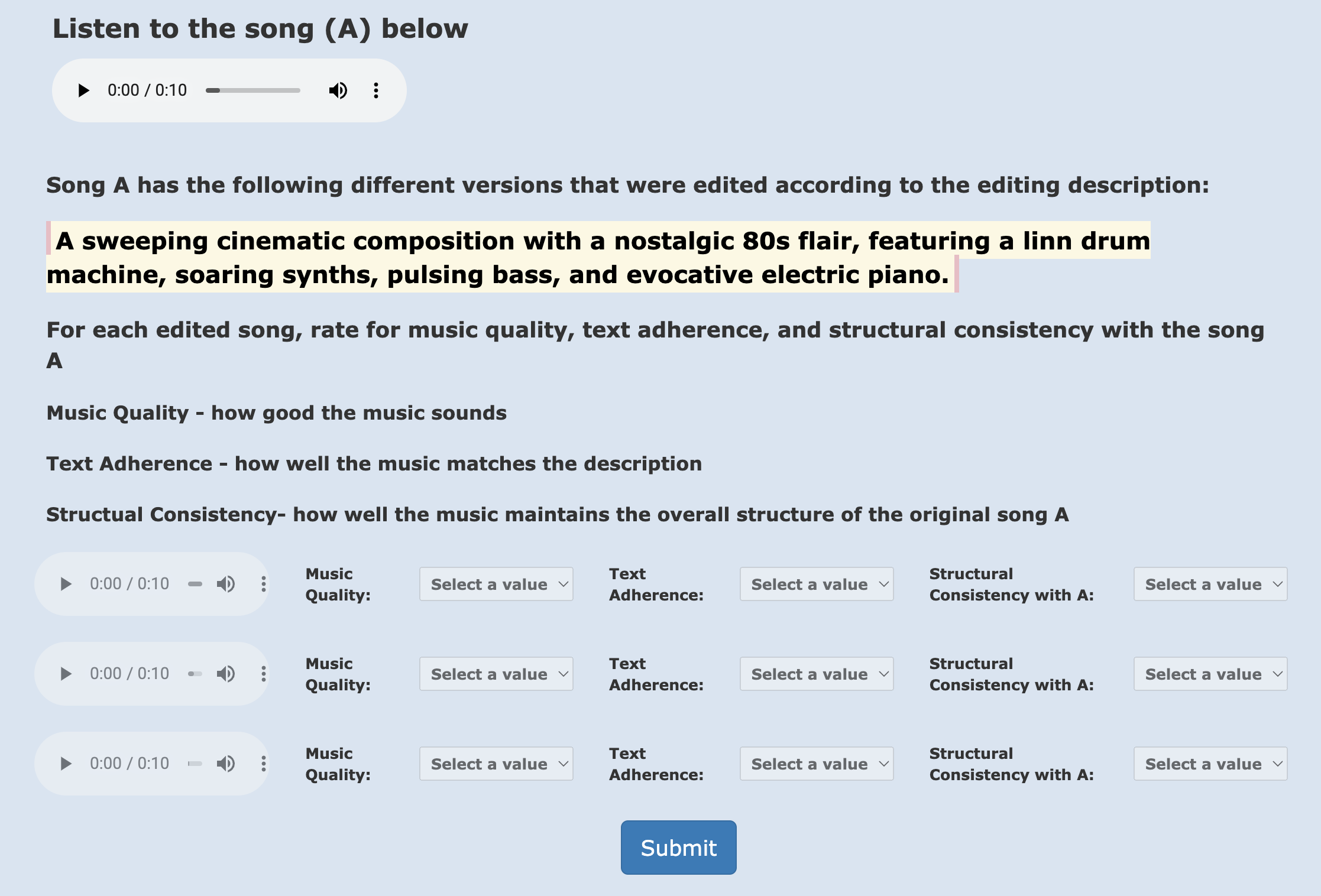}
     % \caption{Generation FAD vs. NFE.}
     \caption{Music editing subjective evaluation form. Given the original song A, raters are asked to evaluate three different edits of A, on the three following axes: quality, text adherence, consistency.}\label{fig:amt_3}
     % \vspace{-0.2cm}
\end{figure}

A screenshot of the music subjective evaluation form is shown in the Figure \ref{fig:amt_3}.
% and \ref{fig:amt_2}.

% \begin{figure}[t!]
    % \vspace{-0.2cm}
%      \centering
%      \includegraphics[width=\textwidth]{figs/amt_1.png}
     % \caption{Generation FAD vs. NFE.}
%      \caption{Part 1 of the edited music subjective evaluation form.}\label{fig:amt_1}
     % \vspace{-0.2cm}
% \end{figure}

% \begin{figure}[t!]
    % \vspace{-0.2cm}
%      \centering
%      \includegraphics[width=\textwidth]{figs/amt_2.png}
     % \caption{Generation FAD vs. NFE.}
%      \caption{Part 2 of the edited music subjective evaluation form.}\label{fig:amt_2}
     % \vspace{-0.2cm}
% \end{figure}

\subsection{Additional experiments}

\subsubsection{Stereo codec}
\label{apx:stereo}

% Given than stereo and sampling rate play an important role in perceived quality and that sound spatialization usually depends on the instruments and genre, we trained \proposed-medium on 30 seconds segments encoded by a 25 Hz 48 Hz stereo codec.
% Aiming for 48 kHz stereo support we train \proposed-medium on 30 seconds segments encoded by a 25 Hz frame rate codec.
The Table \ref{tab:stereo} reports the impact of scaling from mono to stereo with the same \proposed-medium model size (1B parameters) trained on 30 second segments.
Two codecs are trained on 48 kHz audio, using the same 25 Hz latent frame rate: one mono and one stereo.
The in-domain FAD is reported for 10s and 30s generated segments.
Moving from mono to stereo marginally affects the generative model performance.

\subsubsection{Length generalization}
\label{apx:length}

One drawback of training the FM model of a fixed segment duration is that the inference can only be run for the same duration, otherwise the quality will degrade (this can be seen in the FAD$_{10s}$ column of the Table \ref{tab:stereo}, when comparing the first two rows). This can be handled by using padded segments and specific conditioning \citep{evans2024fast}, but does not save any resource when targeting shorter segments.
Another solution is to train on variable length segments but then the model does not generalize well for full length segments, and will better learn for the uttermost left positions of the sequence that appear more often.
We propose to simulate training on variable length segments, while keeping the model learning for the full length scenario.
This is done by applying a L-shaped attention mask during model.
For each sequence of length L, we randomly select a segment boundary within the range $[0, L]$. Positions before the boundary can only attend to themselves in the DiT's self-attention, while positions after it attend to the entire sequence.

% We also report the effect of applying a L-shaped attention mask during model training to better support durations shorter than 30 seconds during inference.
% This is  training on variable length segments, while keeping the model learning for the full length scenario (otherwi

% In order to support generation or editing of samples that may be shorter than 30 seconds we apply a L-shaped attention mask during training only.
% \gael{, schematized in the Figure \ref{fig:lmask}.}
% For each training sequence of length $L$ we uniformly sample a segment boundary in $[0, L]$.
% Positions on the left of the boundary can only attend to themselves in the transformer self-attention while positions on the right always attend to the whole sequence.
% Positions on the left of the boundary can only attend to themselves in the transformer self-attention (hence simulating training on arbitrary duration, shorter than 30 seconds), while positions on the right are allowed to attend to the whole sequence (hence keeping the model ability to learn on full sequences).
% The table \ref{tab:stereo} reports a performance comparison ablating on the L-shaped mask and influence of moving from a mono to stereo codec.
% The in-domain FAD is reported for 10s and 30s generated segments.
Comparing the first two rows of the Table \ref{tab:stereo} indicate that the L-shaped mask helps supporting versatile duration with no penalty on full-length segments, unlocking faster inference for segments shorter than 30 seconds.
This method does not generalize to segments longer than 30 seconds, which should be specifically handled with a sliding window/outpainting approach.
% However moving from mono to stereo marginally affects the generative model performance.
% , with a maintained FAD$_{30s}$ of 0.65, while FAD$_{10s}$ decreases from 1.48 down to 0.59.
% This is especially useful as it enables faster inference whenever a shorter duration is considered.
% On the codec side, moving from mono to stereo only slighlty affects the generative model performance, probably due to the added spatialization information to learn for the same parameter count.

\begin{table}[t!]
  \centering\small
  \caption{Ablation on L-shaped attention mask and stereo for \proposed-large. Each variant is trained on 30s audio segments encoded with a 25 Hz frame rate codec trained on 48 kHz audio.}
  \label{tab:stereo}
  % \resizebox{0.9\columnwidth}{!}{
%   \setlength{\tabcolsep}{3pt}
  \begin{tabular}{c|cc|c|cc}
    \toprule    
    \textsc{Channels} &\textsc{STFT}$_{loss} \downarrow$ &\textsc{SI-SDR}$\uparrow$ &\textsc{L-mask} &FAD$_{10s} \downarrow$ &FAD$_{30s} \downarrow$\\
    \midrule
    \multirow{2}{*}{2} & \multirow{2}{*}{0.40} & \multirow{2}{*}{12.48} & \ding{51} & 0.59 & 0.65 \\
     & && \ding{55} & 1.48 & 0.65 \\
    \midrule
    1 & 0.39 & 13.34 &\ding{51} & 0.49 & 0.59 \\
%     Mono & 20 & 32 & 3.30 & 1.29 & 0.33 \\
%     Stereo & 25 & 48 & 3.78 & 1.31 & 0.31 \\
%     Stereo & 50 & 48 & 3.45 & 1.25 & 0.33 \\
    \bottomrule
  \end{tabular}%}
  % \vspace{-0.4cm}
\end{table}

\begin{figure}[t!]
    % \vspace{-0.2cm}
     \centering
     % \begin{subfigure}[b]{0.32\textwidth}
     %     \centering
     %     \includegraphics[width=\textwidth]{figs/cfg.pdf}
     %     \caption{CFG ablation.}
     %     \label{fig:thr_bz}
     % \end{subfigure}
     % \hfill
     % \begin{subfigure}[b]{\textwidth}
         % \centering
     % \includegraphics[width=0.5\textwidth]{figs/cfg.pdf}
     % \caption{Text-to-music generation quality (FAD) as a function of the classifier-free guidance factor.}
     % \label{fig:cfg}
     \begin{subfigure}[b]{0.49\textwidth}
         \centering
         \includegraphics[width=\textwidth]{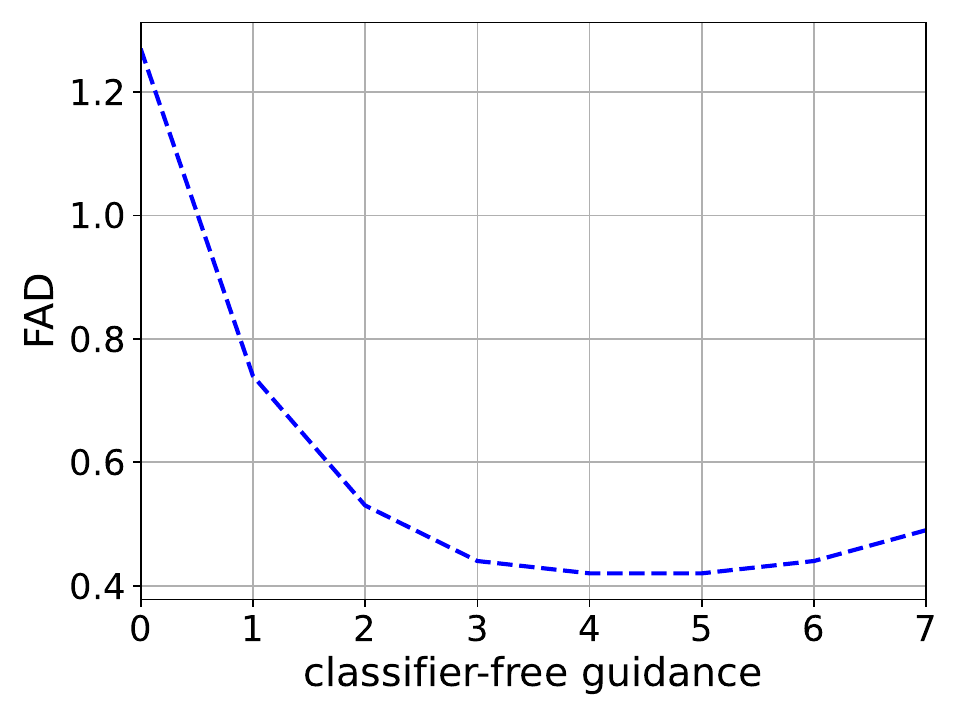}
         \caption{Generation CFG.}
         \label{fig:cfg}
     \end{subfigure}
     \hfill
     \begin{subfigure}[b]{0.49\textwidth}
         \centering
         \includegraphics[width=\textwidth]{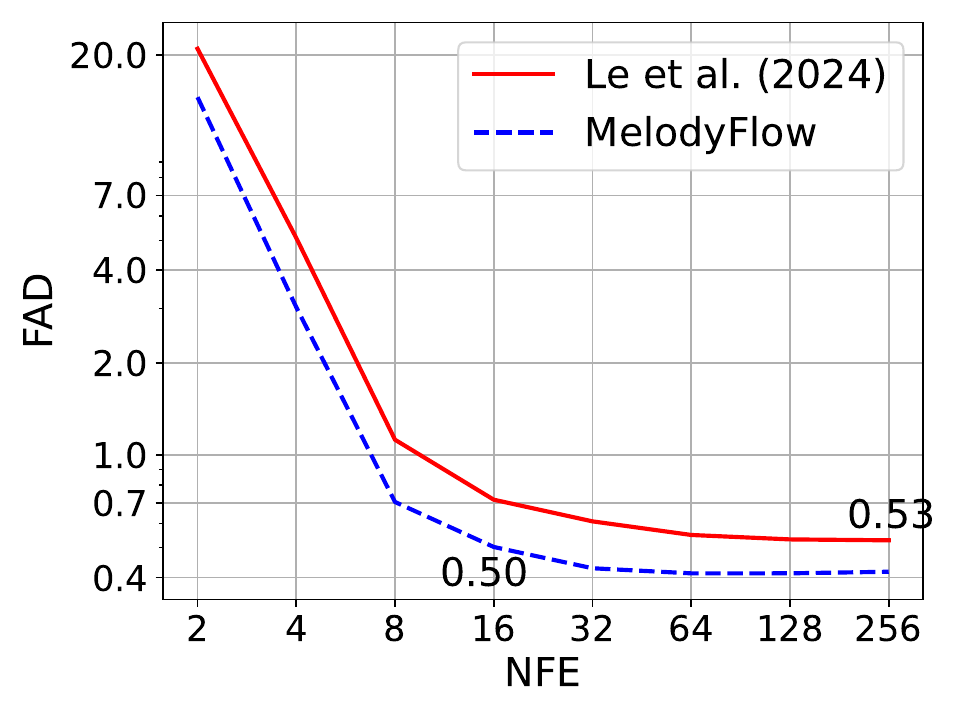}
         \caption{Generation FAD vs. NFE.}
         \label{fig:nfe_gen}
     \end{subfigure}
     %\hfill
     % \end{subfigure}
     % \hfill
     % \begin{subfigure}[b]{0.49\textwidth}
     %     \centering
     %     \includegraphics[width=\textwidth]{figs/lkl_clap.pdf}
     %     \caption{CLAP$_{edit}$ as a function of $\lambda_{KL}$.}
     %     \label{fig:lkl_clap}
     % \end{subfigure}
     % \hfill
     \hfill
     \caption{Text-to-music generation quality (FAD) as a function of classifier-free guidance factor (Figure \ref{fig:cfg}) and inference steps (Figure \ref{fig:nfe_gen}). The baseline of the Figure \ref{fig:nfe_gen} is the FM model architecture of \citep{le2024voicebox} but retrained on our music latents. The combination of our flow matching design choices enable faster generation for a given efficiency budget or better overall quality.}
     % \vspace{-0.2cm}
\end{figure}

\subsubsection{Text-to-music generation efficiency}
\label{apx:efficiency}
% Inference efficiency is quantified by the number of function evaluation (NFE), which denotes the number of
% time an ODE solver evaluates the derivative.
% With diffusion models the generation or editing quality depends on the chosen solver and the number of function evaluation (NFE) - or inference steps - which denotes the number of time an ODE solver estimates the velocity.
In the Figure \ref{fig:nfe_gen} we report the text-to-music generation test FAD as a function of the number of DiT forward passes (NFEs) for both the baseline FM architecture (\citet{le2024voicebox}) and final version of \proposed. Not only does \proposed achieve better performance, but with 16 times fewer NFEs (e.g. where the baseline required 256 NFEs to reach 0.53 FAD, \proposed only requires 16 NFEs to score 0.50).

\begin{figure}[t!]
    % \vspace{-0.2cm}
     \centering
     % \begin{subfigure}[b]{0.32\textwidth}
     %     \centering
     %     \includegraphics[width=\textwidth]{figs/cfg.pdf}
     %     \caption{CFG ablation.}
     %     \label{fig:thr_bz}
     % \end{subfigure}
     % \hfill
     % \begin{subfigure}[b]{\textwidth}
         % \centering
     % \includegraphics[width=0.5\textwidth]{figs/cfg.pdf}
     % \caption{Text-to-music generation quality (FAD) as a function of the classifier-free guidance factor.}
     % \label{fig:cfg}
     \begin{subfigure}[b]{0.32\textwidth}
         \centering
         \includegraphics[width=\textwidth]{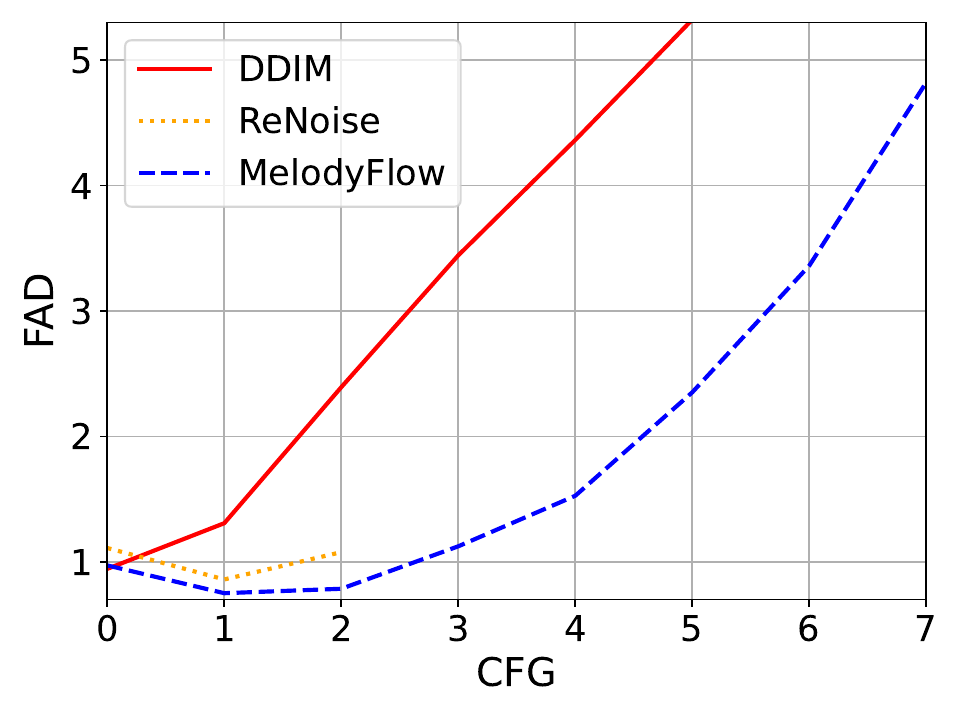}
         \caption{Editing FAD vs. CFG.}
         \label{fig:cfg_edit_fad}
     \end{subfigure}
     \hfill
     \begin{subfigure}[b]{0.32\textwidth}
         \centering
         \includegraphics[width=\textwidth]{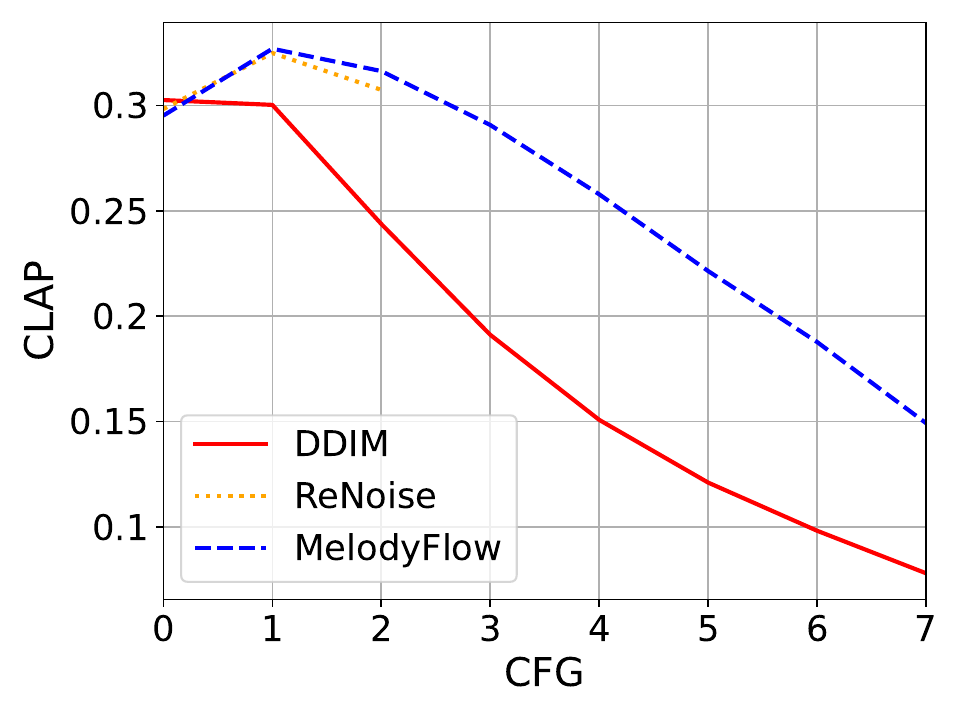}
         \caption{Editing CLAP vs. CFG.}
         \label{fig:cfg_edit_clap}
     \end{subfigure}
     \hfill
     \begin{subfigure}[b]{0.32\textwidth}
         \centering
         \includegraphics[width=\textwidth]{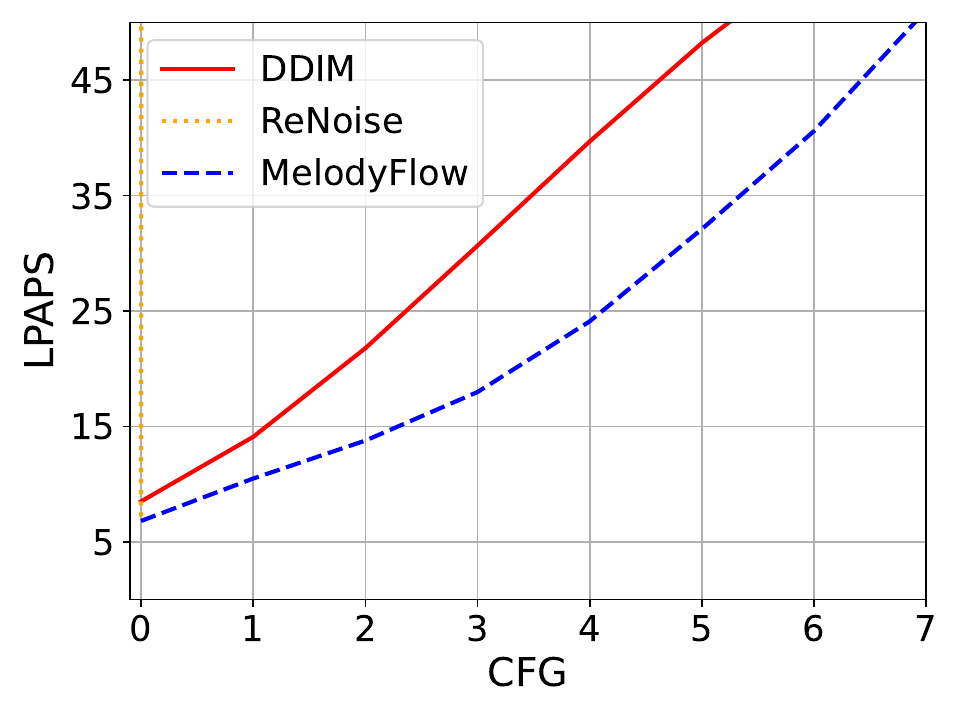}
         \caption{Editing LPAPS vs. CFG.}
         \label{fig:cfg_edit_mse}
     \end{subfigure}
     %\hfill
     % \end{subfigure}
     % \hfill
     % \begin{subfigure}[b]{0.49\textwidth}
     %     \centering
     %     \includegraphics[width=\textwidth]{figs/lkl_clap.pdf}
     %     \caption{CLAP$_{edit}$ as a function of $\lambda_{KL}$.}
     %     \label{fig:lkl_clap}
     % \end{subfigure}
     % \hfill
     \hfill
     \caption{Music editing objective metrics as a function of the classifier free guidance, using the same CFG for both inversion and regeneration.}
     % \vspace{-0.2cm}
\end{figure}

\begin{figure}[t!]
    % \vspace{-0.2cm}
     \centering
     % \begin{subfigure}[b]{0.32\textwidth}
     %     \centering
     %     \includegraphics[width=\textwidth]{figs/cfg.pdf}
     %     \caption{CFG ablation.}
     %     \label{fig:thr_bz}
     % \end{subfigure}
     % \hfill
     % \begin{subfigure}[b]{\textwidth}
         % \centering
     % \includegraphics[width=0.5\textwidth]{figs/cfg.pdf}
     % \caption{Text-to-music generation quality (FAD) as a function of the classifier-free guidance factor.}
     % \label{fig:cfg}
     \begin{subfigure}[b]{0.32\textwidth}
         \centering
         \includegraphics[width=\textwidth]{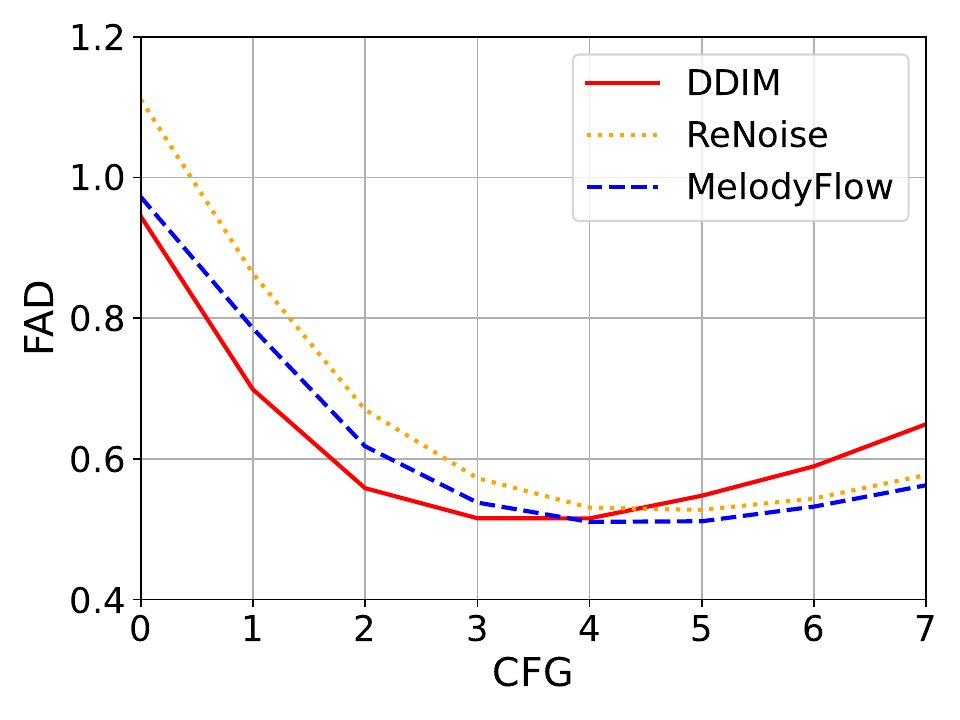}
         \caption{Editing FAD vs. CFG.}
         \label{fig:unc_edit_fad}
     \end{subfigure}
     \hfill
     \begin{subfigure}[b]{0.32\textwidth}
         \centering
         \includegraphics[width=\textwidth]{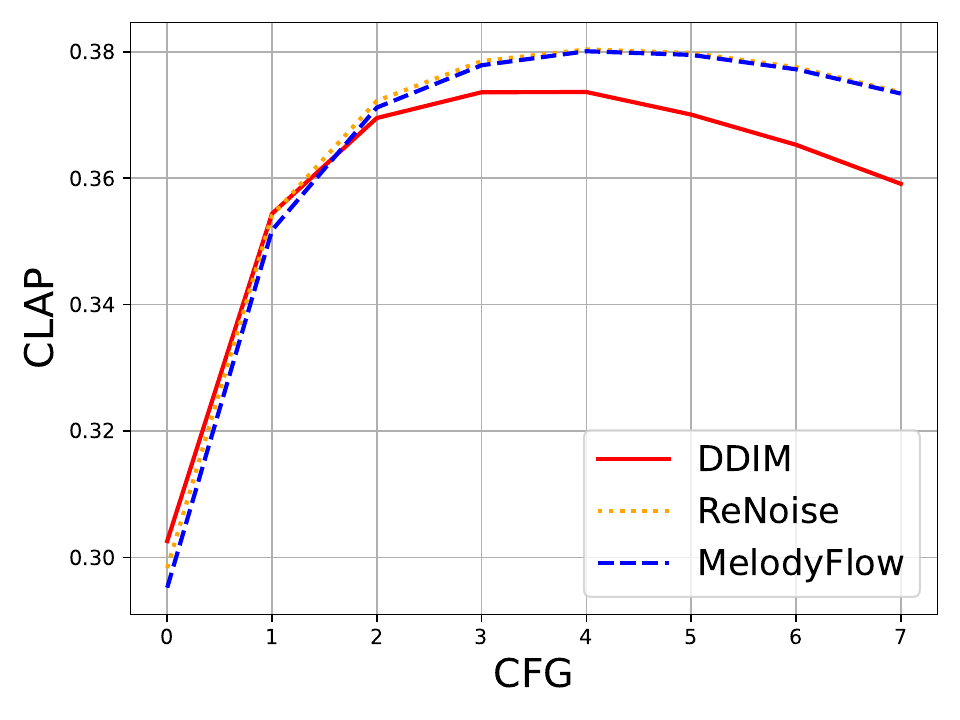}
         \caption{Editing CLAP vs. CFG.}
         \label{fig:unc_edit_clap}
     \end{subfigure}
     \hfill
     \begin{subfigure}[b]{0.32\textwidth}
         \centering
         \includegraphics[width=\textwidth]{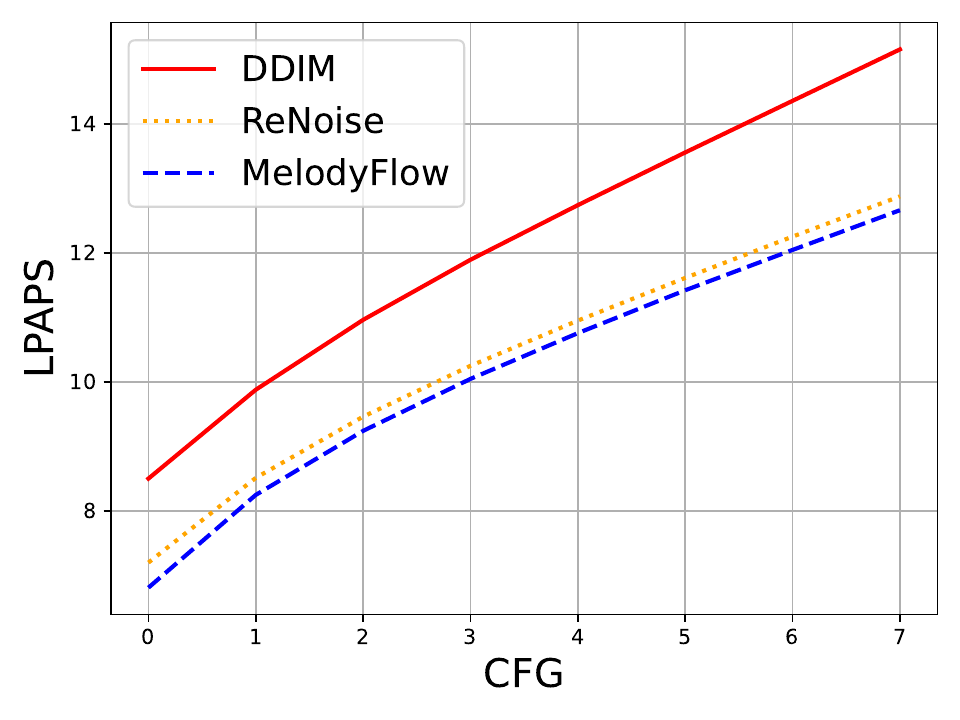}
         \caption{Editing LPAPS vs. CFG.}
         \label{fig:unc_edit_mse}
     \end{subfigure}
     %\hfill
     % \end{subfigure}
     % \hfill
     % \begin{subfigure}[b]{0.49\textwidth}
     %     \centering
     %     \includegraphics[width=\textwidth]{figs/lkl_clap.pdf}
     %     \caption{CLAP$_{edit}$ as a function of $\lambda_{KL}$.}
     %     \label{fig:lkl_clap}
     % \end{subfigure}
     % \hfill
     \hfill
     \caption{Music editing objective metrics as a function of the classifier free guidance, when using a CFG of 0 during latent inversion.}
     % \vspace{-0.2cm}
\end{figure}

\subsubsection{Classifier-free guidance}
\label{apx:cfg}

In the Figure \ref{fig:cfg} we report the in-domain test FAD as a function of the classifier-free guidance factor in the text-to-music generation setting.
We use a classifier-free guidance factor of $4.0$ for the text-to-music generation inference.

In the Figures \ref{fig:cfg_edit_fad}, \ref{fig:cfg_edit_clap}, \ref{fig:cfg_edit_mse} we plot our objective metrics for text-guided music editing, as a function of the CFG.
The performance is bad whatever the considered inversion method, showing that using the CFG during inversion is detrimental.
Above a CFG of 0 \renoise completely diverges (the LPAPS skyrockets), while \proposed achieves the best robustness.
This explains why we consider that \renoise is not directly compatible with the FM formulation, even after converting to $\epsilon$-prediction, and that FM requires an adapted regularized latent inversion method.
After keeping the CFG to zero during latent inversion to stabilize the process, the results are presented in the Figures \ref{fig:unc_edit_fad}, \ref{fig:unc_edit_clap}, \ref{fig:unc_edit_mse}.
We end up using the same classifier-free guidance factor of $4.0$ for our editing experiments.

\end{document}